%% file: ex_article.tex
\documentclass[onefignum,onetabnum]{siamonline220329}

\input{ex_shared}

\ifpdf
\hypersetup{
  pdftitle={Can the clocks tick together despite the noise? Stochastic simulations and analysis},
  pdfauthor={S. M. C. Abo, J. A. Carrillo, and A. T. Layton}
}
\fi

\newcommand{\review}[1]{\textcolor{black}{#1}}
\newcommand{\reviewtwo}[1]{\textcolor{black}{#1}}
\newcommand{\authors}[1]{\textcolor{black}{#1}}

\begin{document}

\maketitle

% REQUIRED
\begin{abstract}
  The suprachiasmatic nucleus (SCN), also known as the circadian master clock, consists of a large population of oscillator neurons. Together, these neurons produce a coherent signal that drives the body's circadian rhythms. What properties of the cell-to-cell communication allow the synchronization of these neurons, despite a wide range of environmental challenges such as fluctuations in photoperiods? To answer that question, we present a mean-field description of globally coupled neurons modeled as Goodwin oscillators with standard Gaussian noise. Provided that the initial conditions of all neurons are independent and identically distributed, any finite number of neurons becomes independent and has the same probability distribution in the mean-field limit, a phenomenon called propagation of chaos. This probability distribution is a solution to a Vlasov-Fokker-Planck type equation, which can be obtained from the stochastic particle model. We study, using the macroscopic description, how the interaction between external noise and intercellular coupling affects the dynamics of the collective rhythm, and we provide a numerical description of the bifurcations resulting from the noise-induced transitions. Our numerical simulations show a noise-induced rhythm generation at low noise intensities, while the SCN clock is arrhythmic in the high noise setting. Notably, coupling induces resonance-like behavior at low noise intensities, and varying coupling strength can cause period locking and variance dissipation even in the presence of noise.
\end{abstract}

% REQUIRED

% REQUIRED

%-----------------------------------------
\section{Introduction}
%-----------------------------------------
The suprachiasmatic nucleus (SCN) in the brain serves as the central clock in mammals and regulates most circadian rhythms in the body \cite{hastings2018generation, refinetti2019circadian}. The SCN is remarkable -- it not only synchronizes the biological rhythms to the external light–dark cycle \cite{abraham2010coupling}, but also generates robust rhythmic outputs with an endogenous period of around 24 h in constant darkness \cite{daido2001circadian,li2020noise}.The specific mechanism responsible for this behaviour continues to be the subject of numerous experimental and theoretical studies. The rhythmic output emanates from a regulatory circuit with a negative feedback loop. We refer to the reviews \cite{partch2014molecular,honma2018mammalian} for a description of the architecture of the SCN clock. 

Although single neurons produce autonomous oscillations, the emergence of global and robust oscillations of the SCN activity requires the synchronization of neural cells \cite{gonze2006circadian}. Oscillations at the global level arise from the interaction, also called \textit{coupling}, between SCN neurons. In this work, we study how a population of SCN neurons manages to synchronize and remain synchronized despite external perturbations. Our focus is on the effect of coupling strength and external noise on synchronization dynamics. Experimental studies have shown that cell-to-cell coupling in the SCN is carried out in part by neurotransmitters \cite{komin2011synchronization,honma2018mammalian}. Vasoactive intestinal polypeptide (VIP), arginine vasopressin (AVP) and gamma-aminobutyric acid (GABA) are examples of neurotransmitters which play a role in the coupling \cite{gu2016synchronization}. \reviewtwo{The SCN is divided into two hemispheres, each of which contains two groups of neurons: a dorsomedial shell (DM) and a ventrolateral (VL) core. These two sets of neurons differ by their light sensitivity, the neurotransmitters they produce, hence their coupling properties. DM cells mainly express AVP, whereas VL neurons express VIP \cite{albus2005gabaergic,hafner2012effect}. Yet, all SCN neurons express the neurotransmitter GABA \cite{moore2002suprachiasmatic}.} In addition to coupling, SCN function is influenced by stochastic noise, which includes exogenous and endogenous cellular noise \cite{st2015quantifying}. Exogenous noise results from changes in the environment \cite{li2020noise,st2015quantifying}, such as fluctuations in light signals, and has been shown to play an important role in the amplitudes of neural oscillators and the entrainment to a new environmental cycle \cite{gu2015noise,fonken2013dim}. The endogenous noise is caused by low molecular counts of the mRNA and protein species involved \cite{st2015quantifying}. 

A number of mathematical models of coupled oscillators have been developed to study the SCN properties such synchrony, the ensemble period and the entrainment ability of the SCN \cite{komin2011synchronization,gonze2005spontaneous,nguyen2013synchronization,pittendrigh1976functional,bernard2007synchronization,gu2009free,locke2008global,gu2011mechanism}. Most of these models are in the form of coupled ordinary differential equations, and are therefore deterministic. Some recent \review{modeling and experimental studies}, however, investigate the influence of noise \review{\cite{gu2015noise,li2020noise,ko2010emergence}} on the circadian clock by means of stochastic differential equations \review{or experimental analysis of stochastic rhythms}. All of these models based on the particle-like description of a set of interacting neurons are called individual-based models (IBM), and often used in animal swarming \cite{ME99,carrillo2010particle}. The topologies often considered are all-to-all coupling between the neurons \cite{komin2011synchronization,gonze2005spontaneous,nguyen2013synchronization} and small world networks \cite{webb2012weakly,vsimonka2015stochastic}. For a large number of interacting agents, the collective motion in the system can be studied through macroscopic descriptions based on the evolution of a density of individuals. These models are known as continuum models, and the scaling limit is called the \textit{mean-field limit} \cite{carrillo2010particle,bolley2011stochastic,Ja}. These continuum models are useful in reducing IBMs into an effective one-body problem: the particle probability density \cite{bolley2011stochastic}. 

Naturally, noise at the level of the IBMs which represent the SCN network is essential since the neuronal activity is not totally deterministic. The randomness should be reflected in the macroscopic description. As pointed out in \cite{carrillo2009double}, stochastic IBMs lead to Fokker-Planck type equations in the mean-field limit for second order models. The proof of this stochastic mean-field limit relies on standard hypotheses: global Lipschitz continuity and linear growth condition of the drift and diffusion coefficients, and the Lipschitz continuity of the interaction function \cite{mckean1967propagation,meleard1996asymptotic,sznitman1991topics}.

In the present article, we consider a stochastic system of interacting SCN neurons in a diffusive scaling and study the effects of external noise, as the network size approaches infinity, on SCN properties: robust oscillation amplitude and period. We also investigate the effect of noise on bifurcation boundaries. \reviewtwo{SCN neurons are characterized by small size and high density \cite{bernard2007synchronization}, and all express GABA \cite{moore2002suprachiasmatic}. We assume, based on this information, that intercellular coupling is carried out by chemical signals released by each cell and that spatial transmission is fast in comparison to the time scale of the oscillations (24h).} We derive the mean-field equation for a system of globally-coupled Goodwin-type neurons with noise. The Goodwin model is commonly employed for circadian oscillators because it describes a biological process with a negative feedback loop --- one of the key circadian clock regulation mechanisms \cite{nguyen2013synchronization}. \reviewtwo{Many studies have considered the SCN as a network where neurons are globally connected \cite{ko2006molecular,komin2011synchronization,gonze2005spontaneous,locke2008global,gu2009free}, but other network topologies for coupling oscillators have also been studied: Newman-Watts (NW) small-world networks \cite{hafner2012effect,vasalou2009small}, regular networks \cite{bernard2007synchronization,kunz2003simulation}, random networks \cite{gu2016circadian} and scale-free networks \cite{gu2016circadian,hafner2012effect}}.

To the best of our knowledge, no study has discussed the influence of external noise on the circadian clock through mean-field equations. We present numerical results on the relation between amplitude of circadian oscillations and coupling strength. We also discuss the effect of noise on bifurcation boundaries. The question arises as to whether fluctuations in the noise level can influence bifurcation boundaries and therefore influence the robustness of circadian oscillations with respect to external noise. Moreover, synchronization will be understood to mean the dissipation of the empirical variance. This approach does not rely on the stability properties of individual neurons nor on the existence of limiting oscillatory behaviors. \cite{bossy2019synchronization}.

The work is organized as follows: in Section \ref{sec:model}, the mean-field model is introduced to describe a network of coupled SCN neurons with noise. Simulation results about the dependence of the rhythms on the coupling strength and noise intensity are discussed in Section \ref{sec:results}. Then, we assess the accuracy of our numerical scheme in Section \ref{sec:validation}. The conclusions and discussion are presented in Section \ref{sec:discussion}. A complete description of the numerical scheme is available in Appendix \ref{app:method}.

%-----------------------------------------
\section{A minimal SCN model and its mean-field description} \label{sec:model}
%-----------------------------------------
\subsection{Model description}
\begin{figure}[htbp]
    \centering
    \includegraphics[width=0.75\linewidth]{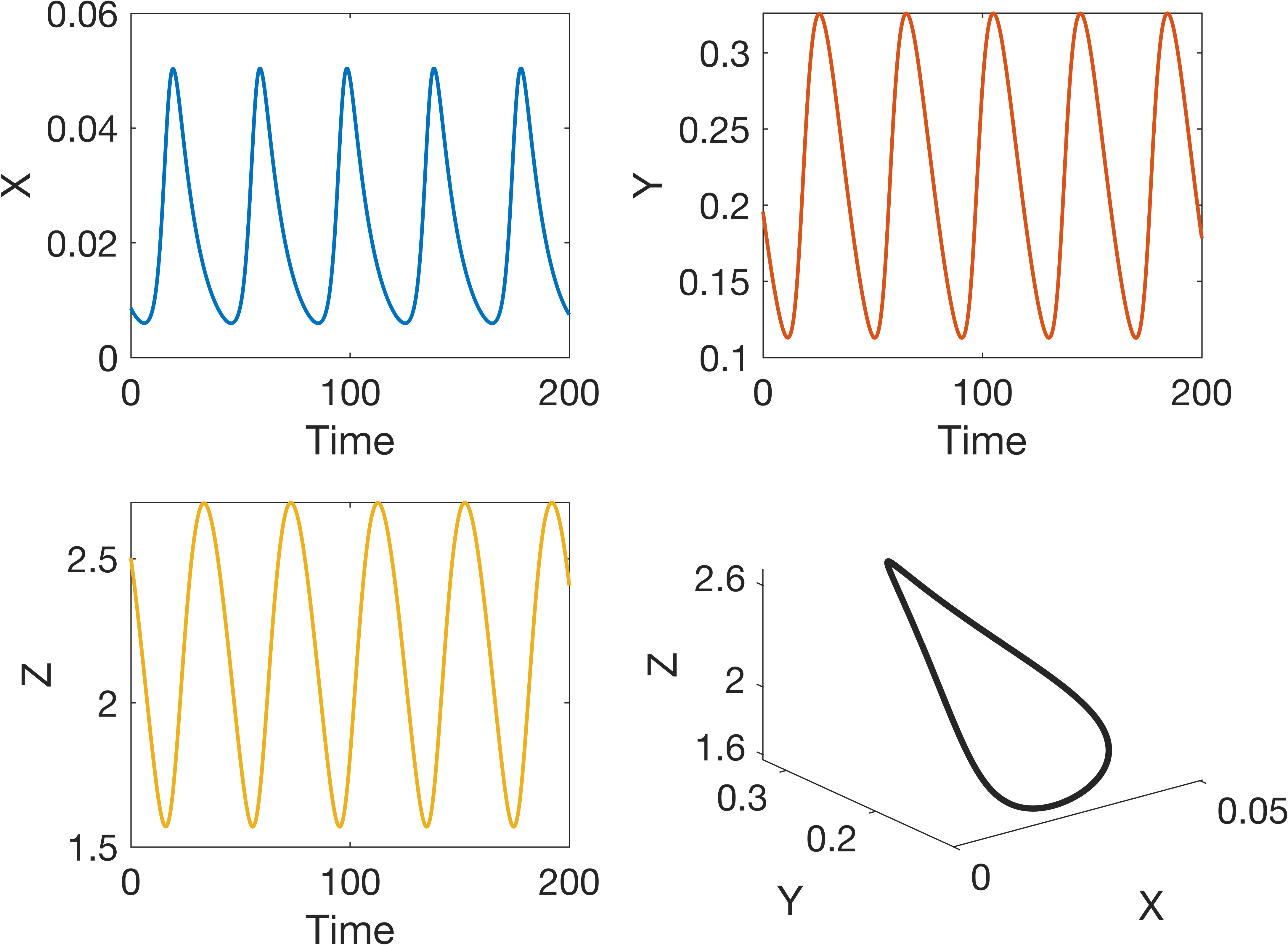}
    \caption{Limit cycle oscillations for the following parameter values: $k_1=1$nM$\cdot$ h$^{-1}$, $k_3=k_5=1$h$^{-1}$, $k_2=k_4=k_6=0.1$h$^{-1}$, $K_i=1$nM, and $n=10$ in (\ref{eq:goodwin}). The oscillation period is about 40h.}
    \label{fig:goodwin-struct}
\end{figure}

In this paper, we propose a mathematical model for describing the collective activity of SCN neurons. The core architecture in mammals of the circadian clock consists of two feedback loops that interact to generate biochemical oscillations with a period of nearly 24 h \cite{relogio2011tuning}. The primary feedback loop is driven by clock proteins CLOCK and BMAL1. The proteins dimerize to create the CLOCK-BMAL1 complex which initiates the transcription of the target period (PER) and cryptochrome (CRY) genes. Negative feedback is achieved through PER-CRY heterodimers that repress their own transcription after delays due to cellular processes, such as transcription, translation, and nuclear transport \cite{ananthasubramaniam2020amplitude}. In a secondary loop, CLOCK-BMAL1 proteins activate the transcription of Rev-Erb$\alpha$.
After being translated into proteins, Rev-Erb$\alpha$ downregulates \textit{Bmal1} transcription, thus completing the loop \cite{relogio2011tuning, ko2006molecular}. The Goodwin model, a negative feedback oscillator with variables $X$, $Y$ and $Z$, can be used to represent these two regulatory loops; see (\ref{eq:goodwin}-\ref{eq:repression}) and Fig~\ref{fig:goodwin-struct}. In general, the mechanism behind biological oscillators consists of delayed negative feedback loops. \cite{tiana2007oscillations}.

The Goodwin model has been widely studied theoretically \cite{woller2014goodwin,ananthasubramaniam2020amplitude,gonze2013goodwin} and applied to various biological systems, such as circadian clocks \cite{komin2011synchronization,gonze2005spontaneous,abraham2010coupling,nguyen2013synchronization} or enzymatic regulation \cite{goodwin1965oscillatory}. The temporal evolution of a single Goodwin-type neuron is governed by the following equations:
\begin{equation}
    \Dot{X} = f(Z) - k_2X,\quad \Dot{Y} = k_3X - k_4Y,\quad \Dot{Z} = k_5Y - k_6Z, 
    \label{eq:goodwin}
\end{equation}
where
\begin{equation}
    f(Z) = k_1 \frac{K_i^n}{K_i^n + Z^n}.
    \label{eq:repression}
\end{equation}

\begin{figure}[htbp]
\centering
\begin{subfigure}{.48\textwidth}
  \centering
  \includegraphics[width=\linewidth]{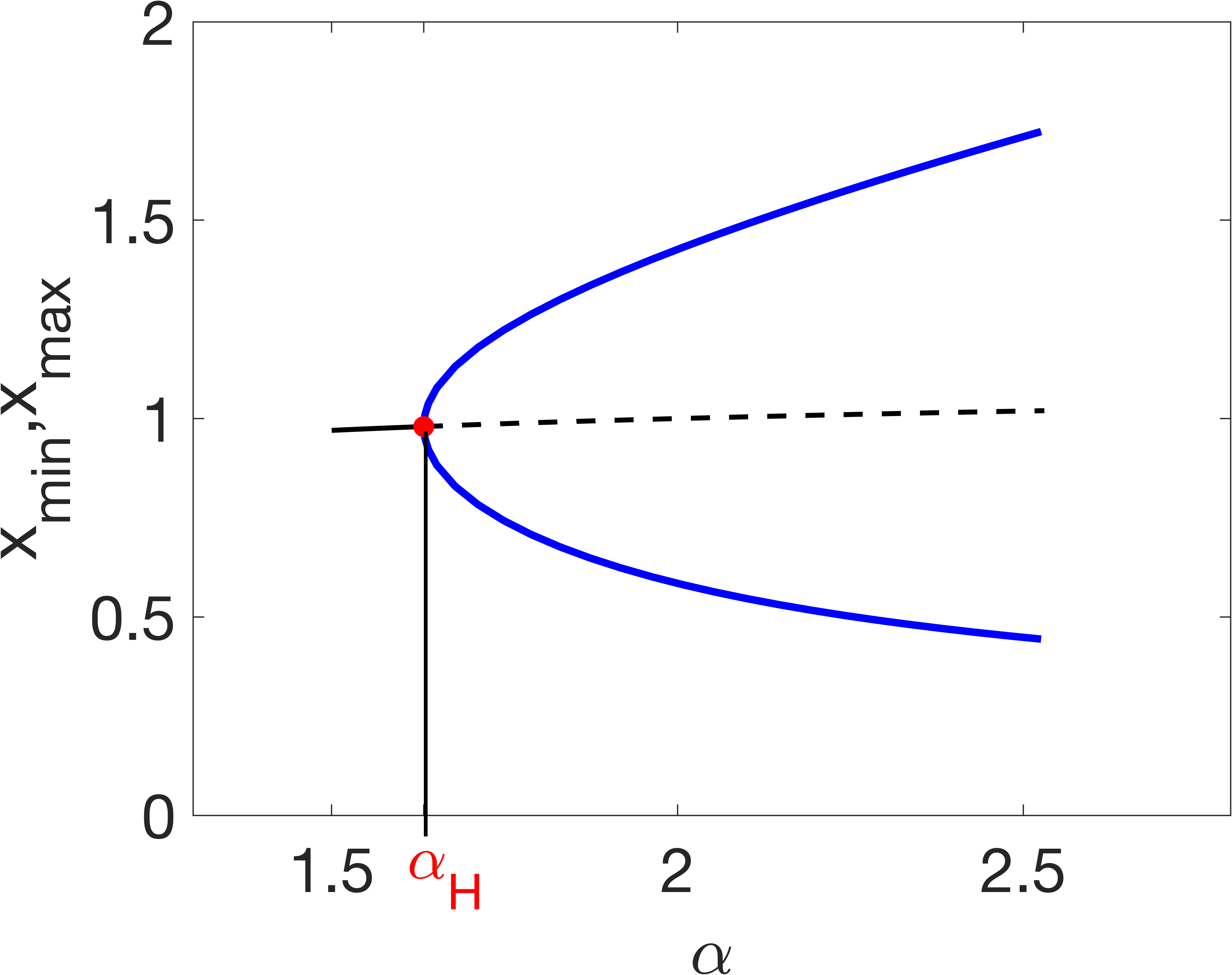} 
  \caption{}
  \label{fig:bifGneuron}
\end{subfigure}
\hspace{0.1cm}
\begin{subfigure}{.48\textwidth}
  \centering
  \includegraphics[width=\linewidth]{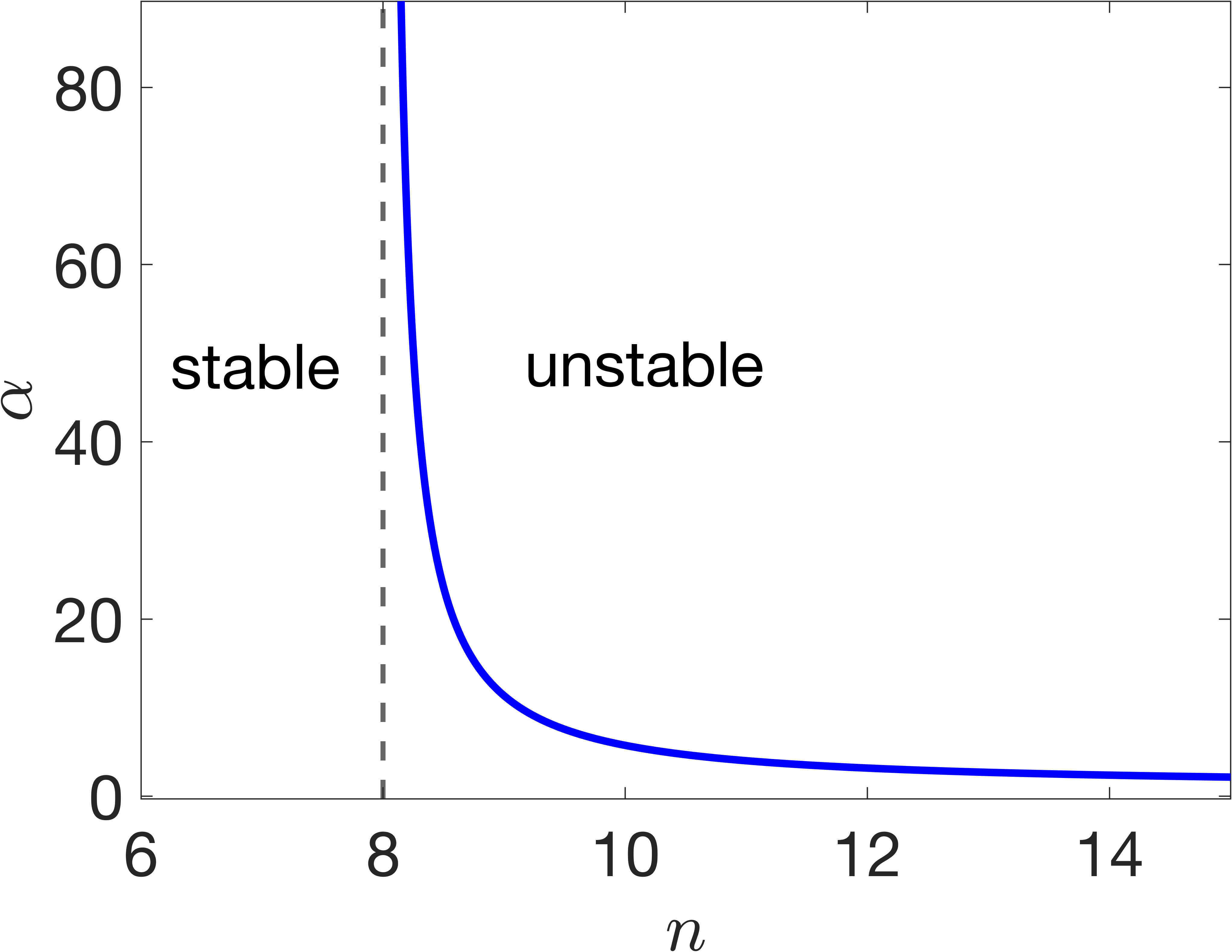}
  \caption{}
  \label{fig:twopar}
\end{subfigure}
\caption{\reviewtwo{(a) Bifurcation diagram of the stable periodic solutions near the Hopf bifurcation point ($\alpha_H=1.633$) when $n=20$. At $\alpha_H$, the system's stability switches from stable (solid black line) to unstable (dashed black line) and a periodic solution arises. The bifurcation is supercritical (solid blue curves). (b) Two-parameter bifurcation diagram in terms of $\alpha$ and $n$. Hill exponent $n>8$ is required for oscillations. If $n<8$, the steady state is a stable focus for all values of $\alpha$.}}
\label{fig:bifGoodwin}
\end{figure}

In this model, $X$ denotes the mRNA concentration of a certain clock gene, $Y$ is the matching protein, and $Z$ is a transcriptional inhibitor in the nuclear form or the phosphorylated form of the protein. The inhibition term is described by a nonlinear and hyperbolic function (i.e. Hill function), $f(Z) $. All other terms are linear. The Hill function is parametrized by a Hill coefficient $n$ characterizing the response steepness, and an inhibition threshold $K_i$ that describes the concentration of inhibitor that halves the production rate, i.e., half-maximal repression occurs when $Z = K_i$. \review{In many organisms, the Hill function has been employed to characterize transcriptional repression: Neurospora \cite{ruoff1996temperature,leloup1999limit}, Drosophila \cite{leloup1998model,ruoff1996temperature,ueda2001robust} and mammals \cite{leloup2003toward,relogio2011tuning,gonze2005spontaneous,komin2011synchronization}. Hill functions are often employed to describe cooperative binding of repressors to the gene promotor in transcription \cite{goldbeter1995model} or repression based on multisite phosphorylation \cite{gonze2013goodwin}. The latter is a more realistic mechanism, especially because a large Hill exponent is required for oscillations in the Goodwin model ($n>8$). The recent work of Cao et al. \cite{cao2021molecular}, which builds upon \cite{robles2017phosphorylation,narasimamurthy2018ck1delta}, provides some evidence for repression based on multisite phosphorylation in mammals. The authors show that removal of CLOCK–BMAL1 involves phosphorylation (hyperphosphorylation) of CLOCK, which is accomplished by CK1$\delta$ when CRY and PER deliver CK1$\delta$ to the CLOCK–BMAL1 complex in the nucleus of cells. A different transcriptional repression mechanism based on protein sequestration has been proposed to describe the negative feedback underlying circadian oscillators. See \cite{kim2012mechanism,kim2014molecular,kim2016protein,chen2021collective} for details.} The various rate constants parametrize transcription $(k_1, k_3)$, degradation $(k_2, k_4, k_6)$, and nuclear import $(k_5)$. Note that all reaction rates are positive. Concentrations $X, Y, Z$ and $K_i$ have units nM. Rate constants have units h$^{-1}$, except for $k_1$ which has units nM h$^{-1}$. Depending on parameter values, the model can produce limit cycle oscillations.
Following the approach in \cite{woller2014goodwin}, we reformulate the equations in (\ref{eq:goodwin}) in dimensionless form. Assuming equal degradation rates ($k_2 = k_4 = k_6$), we introduce the new variables:
\begin{equation}
    x=\frac{k_3k_5}{k_2^2K_i}X, \quad y=\frac{k_5}{k_2K_i}Y, \quad z=\frac{Z}{K_i}, \quad \review{t=\frac{k_2 T}{\tau}}, \nonumber
\end{equation}
with $\tau$ chosen to make the intrinsic period of the oscillator 23.5. We obtain,
\begin{equation}
    \review{\dv{x}{t}} = \frac{\alpha}{1+z^n} - x,\quad \review{\dv{y}{t}} = x-y,\quad \review{\dv{z}{t}} = y-z,
    \label{eq:goodwin-nondim}
\end{equation}
where  
\begin{equation}
    \alpha\equiv\frac{k_1k_3k_5}{k_2^3K_i} \label{eq:alpha}
\end{equation}
is the only parameter for a given $n$. \reviewtwo{Fig~\ref{fig:bifGoodwin} represents the bifurcation and stability diagrams for system \eqref{eq:goodwin-nondim}. At the critical value $\alpha_H$, the system transitions from a stable steady state to an unstable steady state (via a Hopf bifurcation), and a periodic solution arises (Fig~\ref{fig:bifGneuron}). The steady state is stable if $\alpha<\alpha_H$ and unstable otherwise. To obtain limit-cycle oscillations, the Hill coefficient must be larger than 8 (Fig~\ref{fig:twopar})}.

We now consider a population of N identical neurons. Let $x_i(t), y_i(t), z_i(t) \in \mathbb{R}$ denote the concentration of mRNA, protein and inhibitor protein of neuron $i$ at time $t$, respectively. To produce a functional SCN network, there must be reciprocal signalling between neurons. We examine our network under \textit{all-to-all} coupling conditions. Each neuron in the group adjusts its production of mRNA ($x$) by averaging with all the others. We obtain,
\begin{equation}
    \begin{cases}
        \dv{x_i}{t} &= \frac{\alpha}{1+z_i^n} - x_i + K(\Bar{x} - x_i) \\
        \dv{y_i}{t} &= x_i-y_i \\
        \dv{z_i}{t} &= y_i-z_i
        \label{eq:network}
    \end{cases}
\end{equation}

Here, and throughout this paper, the coupling parameter $K$ is assumed to be the same for all oscillators. $\Bar{x}(t)$ is the average value of all individual variables $x_i(t)$ at time $t$:
\begin{equation}
  \Bar{x}(t) = \frac{1}{N}\sum_{i=1}^{N}x_i(t)  
\end{equation}

\subsection{Stochastic extension and mean-field limit} \label{sec:mainmodel}
Our goal is to analyze a system of the type (\ref{eq:network}) with Gaussian noise. Biological clocks, although noisy at the microscopic level due to both external noise (e.g., fluctuations in photoperiods) and inherent stochasticity (e.g., coupling between cells), can be relatively precise at the macroscopic level \cite{asgari2019mathematical, herzog2004temporal}. We represent stochasticity via additive white noise acting through the first variable $x$. Specifically, we will consider a large network of N interacting $\mathbb{R}^{3}$-valued processes $(x_i(t),y_i(t),z_i(t))_{t\geq0}$ with $1\leq i \leq N$ solution of
\begin{equation}
\begin{cases}
    \,dx_i(t) = \bigg[\Big(\frac{\alpha}{1+z_i(t)^n} - x_i(t)\Big) \authors{+} \frac{K}{N}\sum\limits_{j=1}^{N} H(x_i(t)-x_j(t))\bigg] dt+\! \sqrt{2D} \,dW_i(t) \\
    \,dy_i(t) = \big(x_i(t)-y_i(t)\big) \,dt \\  
    \,dz_i(t) = \big(y_i(t)-z_i(t)\big) \,dt 
    \label{eq:stochnetwork} 
\end{cases}
\end{equation}
\authors{where $H(\omega):=-\omega$} and with independent and identically distributed initial data $(x_i^0,y_i^0,z_i^0)$, $1\leq i \leq N$. The processes $(W_i(t))_{t\geq 0}$ with $1\leq i \leq N$ are independent Brownian motions in $\mathbb{R}$ and the noise intensity is $\sqrt{2D}$, with $D>0$. To our knowledge, the stochastic mean-field limit description of a system of coupled Goodwin-type neurons has not yet been considered. 

All neurons have the same distribution on $\mathbb{R}^{3}$ at time $t$ due to the symmetry of the initial configuration and of the evolution \cite{bolley2011stochastic}. For any $t > 0$ the neurons become correlated due to the coupling term $\frac{K}{N}\sum\limits_{j=1}^N H(x_j - x_i)$ in the evolution, though they are independent at $t=0$. However, given the order $1/N$ of the interaction term, it seems reasonable that any fixed number $k$ of these interacting neurons become less correlated as N gets large. This property is called propagation of chaos \cite{bolley2011stochastic}. 

The following assumptions hold for the stochastic model: 1) global Lipschitz continuity of the drift and diffusion terms; 2) linear growth condition of the drift and diffusion terms; 3) Lipschitz continuity of the coupling function. These assumptions imply that the system of stochastic differential equations (\ref{eq:stochnetwork}) is well-posed.

It follows from the general theory of Sznitman \cite{sznitman1991topics} (see also the more recent \cite{malrieu2003convergence,bolley2011stochastic}) that our N interacting processes $(x_i(t),y_i(t),z_i(t))_{t\geq0}$ respectively behave as $N \rightarrow \infty$ like the processes $(\Tilde{x}_i(t),\Tilde{y}_i(t),\Tilde{z}_i(t))_{t\geq0}$, solutions of the kinetic McKean-Vlasov type processes on $\mathbb{R}^{3}$:
\begin{equation}
    \begin{cases}
    &\,d\Tilde{x}_i(t) = \psi(\rho) \,dt + \sqrt{2D} \,dW_i(t) \\
    &\,d\Tilde{y}_i(t) = (\Tilde{x}_i(t)-\Tilde{y}_i(t)) \,dt \\  
    &\,d\Tilde{z}_i(t) = (\Tilde{y}_i(t)-\Tilde{z}_i(t)) \,dt \\
    &\big(\Tilde{x}_i^0,\Tilde{y}_i^0,\Tilde{z}_i^0\big) = \big(x_i^0,y_i^0,z_i^0\big), \quad \rho = \text{law}\big(\Tilde{x}_i(t),\Tilde{y}_i(t),\Tilde{z}_i(t)\big)\\
    &\psi\big(\rho\big) \big(\Tilde{x_i},\Tilde{y_i},\Tilde{z_i},t\big) = \frac{\alpha}{1+\Tilde{z_i}(t)^n} - \Tilde{x_i}(t) + K \big(H \star \rho(\Tilde{x}_i,\Tilde{y}_i,\Tilde{z}_i,t)\big)
    \label{eq:stochnetwork2}   
    \end{cases}
\end{equation}

The Brownian motions $(W_i(t))_{t\geq 0}$ in (\ref{eq:stochnetwork2}) are those governing the evolution of $(x_i(t),y_i(t),\\z_i(t))_{t\geq0}$. The processes $(\Tilde{x}_i(t),\Tilde{y}_i(t),\Tilde{z}_i(t))_{t\geq0}$ with $i \geq 1$ are independent since the initial conditions and governing Brownian motions are independent. Notice that they are identically distributed and, by the Itô formula, their common law $\rho$ at time $t$ should evolve according to the kinetic McKean-Vlasov equation
\begin{equation}
  \pdv{\rho}{t} =  D \partial_x^2 \rho -  \partial_x \big[\xi(\rho)\rho\big] -  \partial_y \big[(x-y) \rho \big] -  \partial_z \big[(y-z)\rho\big]
  \label{eq:mainPDE}
\end{equation}
where 
\begin{equation}
    \xi\big(\rho\big) \big(x,y,z,t\big) = \frac{\alpha}{1+z^n} - x +  K \big(H\star\rho\big) \label{eq:xi}
\end{equation}
with 
\begin{equation}
    H \star \rho(\authors{x},y,z,t) = \int_{\mathbb{R}^{3d}} H(x-w) \rho(w,y,z,t) \,dw\,dy\,dz \,. \label{eq:convol}
\end{equation}

Since (\ref{eq:stochnetwork2}) models the evolution of concentrations $(\Tilde{x}_i(t),\Tilde{y}_i(t),\Tilde{z}_i(t))_{t\geq0}$ we constrained the solutions of the network when performing numerical simulations to ensure that they remained in a smooth positive domain in $\mathbb{R}^{3}_{+}$. In particular, we used the compact support $[0,2]\times[0,2]\times[0,2]$. The boundaries of the domain are instantaneously reflecting in an oblique direction. See \cite{gobet2001euler, hanks2017reflected} and the references therein for a detailed discussion of Euler schemes for reflected stochastic differential equations. In this case, the general theory of Snitzman \cite{Sznitman-1984-nonlinear-reflecting-diffusion,Lions-Sznitman-1984-SDEreflectingBC,sznitman1991topics} still applies, but we recover (\ref{eq:mainPDE}) with no-flux boundary conditions.

%-----------------------------------------
\section{Numerical results} \label{sec:results}
%-----------------------------------------
We focus on the dynamic evolution of solutions in the mean field scaling of the SCN network with noise. Our model (\ref{eq:mainPDE}) is a nonlocal nonlinear transport equation with no-flux boundary conditions. Such characteristics of the mean-field equation make it difficult to conduct a theoretical study using center manifold or bifurcation theory \cite{drogoul2017hopf,drogoul2016exponential}. Instead, we investigate the nature of the bifurcations numerically and compare the solution to the mean-field equation with that of the finite SCN network. We consider two main parameters in our analysis, namely the strength of the coupling $K$ and the noise level $D$. In all our simulations, we refer to the marginal probability densities for $x$, $y$ and $z$ as $\rho_1$, $\rho_2$, and $\rho_3$, respectively.

\reviewtwo{A noise-free network analysis is outside the scope of this article, but we refer the reader to \cite{woller2014goodwin,chen2021collective} for details on the stability of the noise-free system and associated bifurcation diagrams. In the case of the noise free network with identical Goodwin cells \eqref{eq:network}, the linear stability of steady states is tractable and bifurcation diagrams can be determined by solving for the steady states of the amplitude equations \cite{zhang2011periodically}. Moreover, for oscillators with weak coupling the phase-locked states could be studied using weakly coupled oscillator theory, where the network can be reduced to its phase model description \cite{ermentrout1984frequency,Kuramoto1984,neu1979coupled}. The stability of the synchronous state for strong coupling can be studied using the master stability function, which allows to calculate the stability as determined by a particular choice of stability measure, like Lyapunov or Floquet exponents \cite{pecora1998master,sun2009master}.}

\subsection{Coupling strength and synchronization} \label{sec:coupling}

The SCN coordinates physiological cycles throughout the body with incredible precision \cite{herzog2017regulating}. Nevertheless, local oscillations at the level of individual neurons can be substantially different from global network-wide oscillations \cite{herzog2017regulating,schmal2018measuring}. In fact, defining characteristics of circadian rhythms such as period, large amplitude, accuracy and synchrony arise due to coupling \cite{schmal2018measuring,welsh2010suprachiasmatic,mohawk2011cell}. In this section, we investigate the effect of coupling in the overall dynamics of the SCN. We perform a numerical bifurcation analysis by varying the parameter $K$, \review{which quantifies the strength of the chemical signal that is transmitted to oscillator cells. A value of $K<1$ signifies a decay of the chemical signal before it reaches the target cell, while $K=1$ represents the perfect transduction of the chemical signal to the target cell. In particular, $K=0$ implies an uncoupled network where oscillations are localized in individual neuronal cells.} Following the approach done in \cite{drogoul2017hopf, aymard2019mean}, a stationary solution of (\ref{eq:mainPDE}) characterises \textit{asynchronous activity}, a state in which neurons exhibit out-of-phase oscillations. \textit{Synchronized activity} refers to a state where the mean-field is periodic in time.

\begin{figure}[htbp]
\centering
\begin{subfigure}{.48\textwidth}
  \centering
  \includegraphics[width=\linewidth]{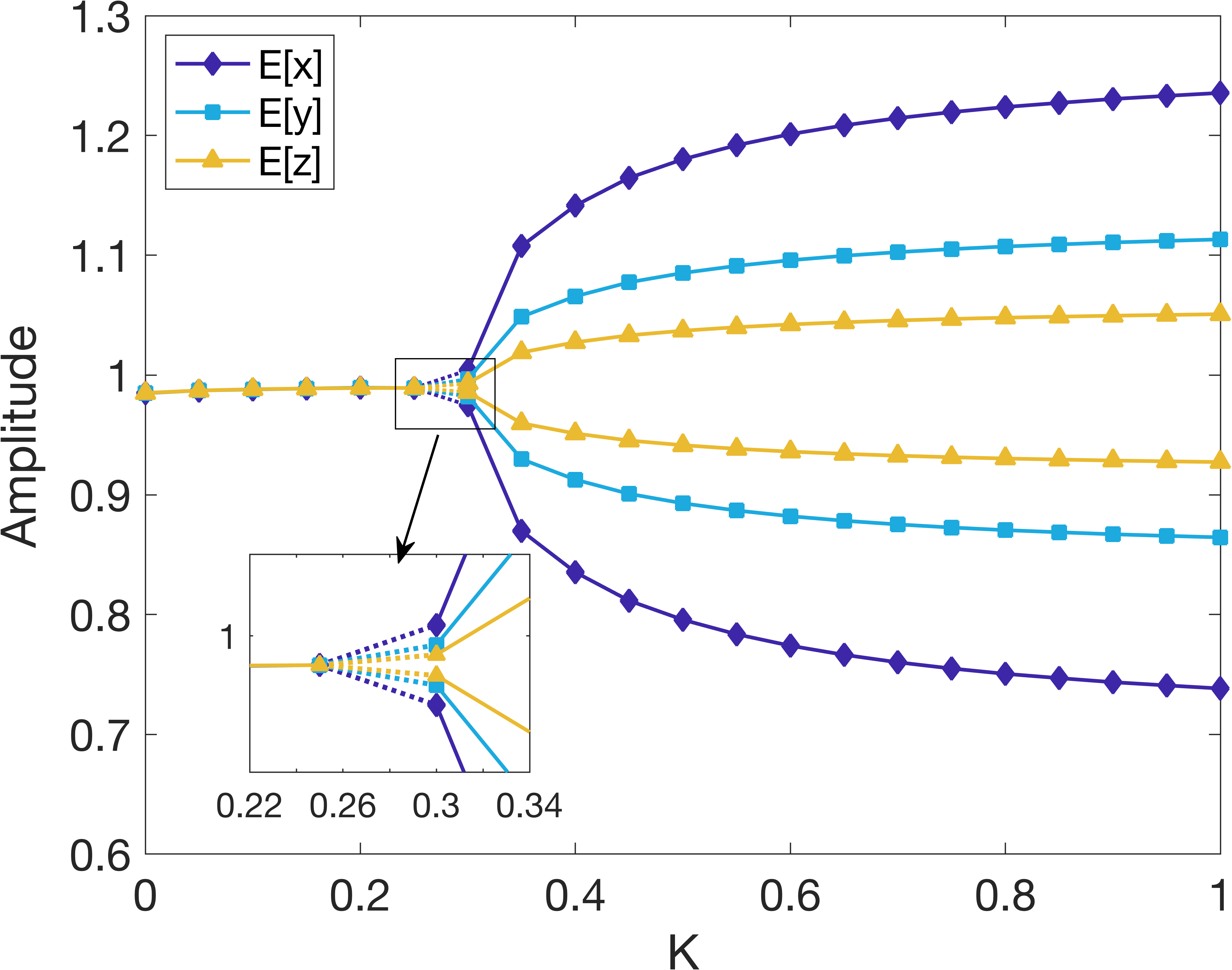} 
  \caption{}
  \label{fig:bif}
\end{subfigure}
\hspace{0.1cm}
\begin{subfigure}{.48\textwidth}
  \centering
  \includegraphics[width=\linewidth]{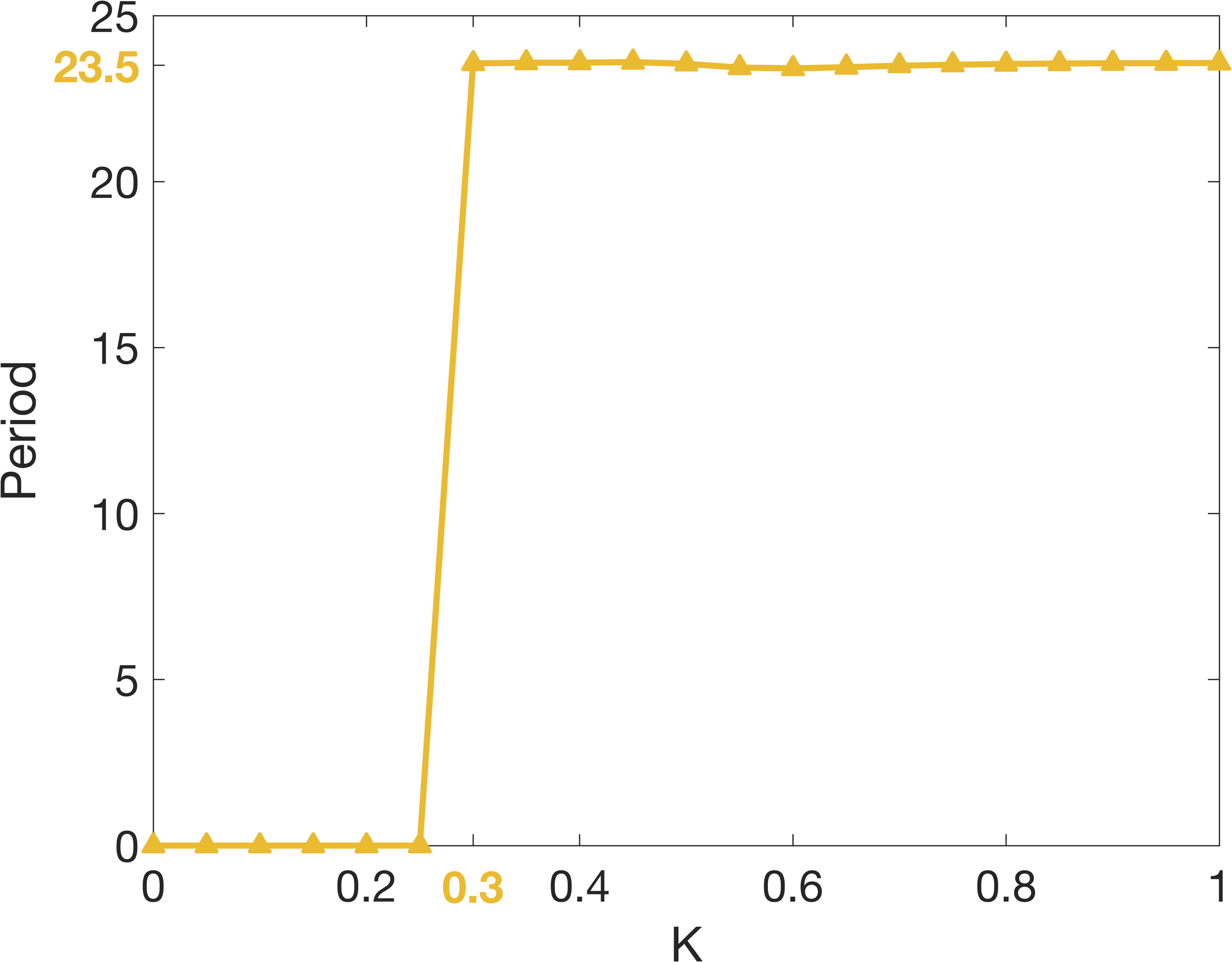} 
  \caption{}
  \label{fig:perK2}
\end{subfigure}
\caption{(a) Bifurcation diagram associated with the coupling strength $K$ in the mean-field model. $E[x]$, $E[y]$ and $E[z]$ refer to spatial averages. Numerical simulations are run until a steady state is reached or until the oscillation amplitude becomes stable. For the latter, the graph shows the peaks and troughs of oscillations as a function of $K$. (b) Period of oscillations of $E[z]$. Noise level is constant at $D=0.01$ in both (a) and (b). A Hopf bifurcation appears around the critical value $K_H \approx 0.3$.}
\label{fig:bifK2}
\end{figure}

In Fig~\ref{fig:bifK2} we show a bifurcation diagram of the spatial averages in $x$, $y$, and $z$ as a function of the coupling strength $K$. $E[x]$, $E[y]$ and $E[z]$ are calculated from the solution to the mean-field equation (\ref{eq:mainPDE}). \review{We call $E[\cdot]$ the average over the values of a given variable across the space domain}. \reviewtwo{We assume a low level of noise with $D=0.01$}. Fig~\ref{fig:bif} shows the transition from a stable regime without oscillations to a regime of sustained oscillations, which correspond to the progression of solutions toward a periodic orbit \cite{gonze2002robustness}. For values of $K$ less than $0.3$, the mean-field solution shows damped oscillations that eventually result in an invariant distribution. This indicates that the network of SCN neurons is out of sync. However, past the critical value $K_H \approx 0.3$, synchronized activity emerges within the network as shown by a periodic solution to (\ref{eq:mainPDE}). Figs~\ref{fig:bif} and \ref{fig:perK2} suggest the existence of a bifurcation of the asynchronous state. At this bifurcation emerges a limit cycle corresponding to a stable and robust sinusoidal oscillation. Mathematically, the expansion of amplitude as the periodic orbit moves away from the bifurcation boundary (Fig~\ref{fig:bif}) and the relatively constant period of the oscillations (Fig~\ref{fig:perK2}) are characteristic of a supercritical Hopf bifurcation. Moreover, the amplitude grows as the coupling strength increases, and this phenomena is especially prominent near the bifurcation boundary (Fig~\ref{fig:bif}). Previous studies have shown that coupling can induce amplitude expansion near the Hopf bifurcation \cite{abraham2010coupling,schmal2018measuring,ashwin2016mathematical}. This so-called ``resonance'' may be enhanced by synchronizing factors, here represented by the coupling strength $K$.

\begin{figure}[htbp]
\begin{subfigure}{.48\textwidth}
  \centering
  \includegraphics[width=\linewidth]{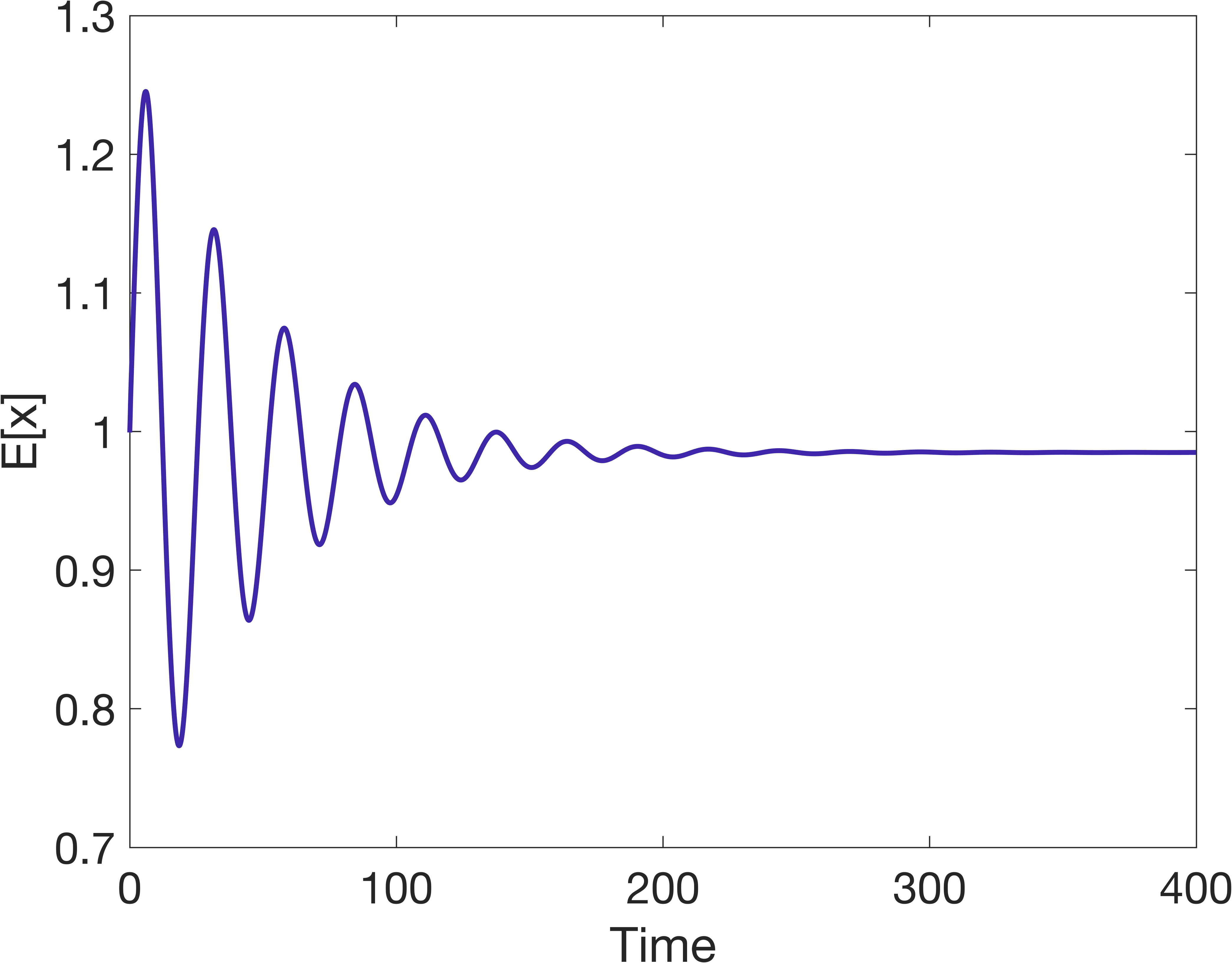} 
  \caption{}
  \label{fig:tsk0}
\end{subfigure}
\hspace{0.1cm}
\begin{subfigure}{.48\textwidth}
  \centering
  \includegraphics[width=\linewidth]{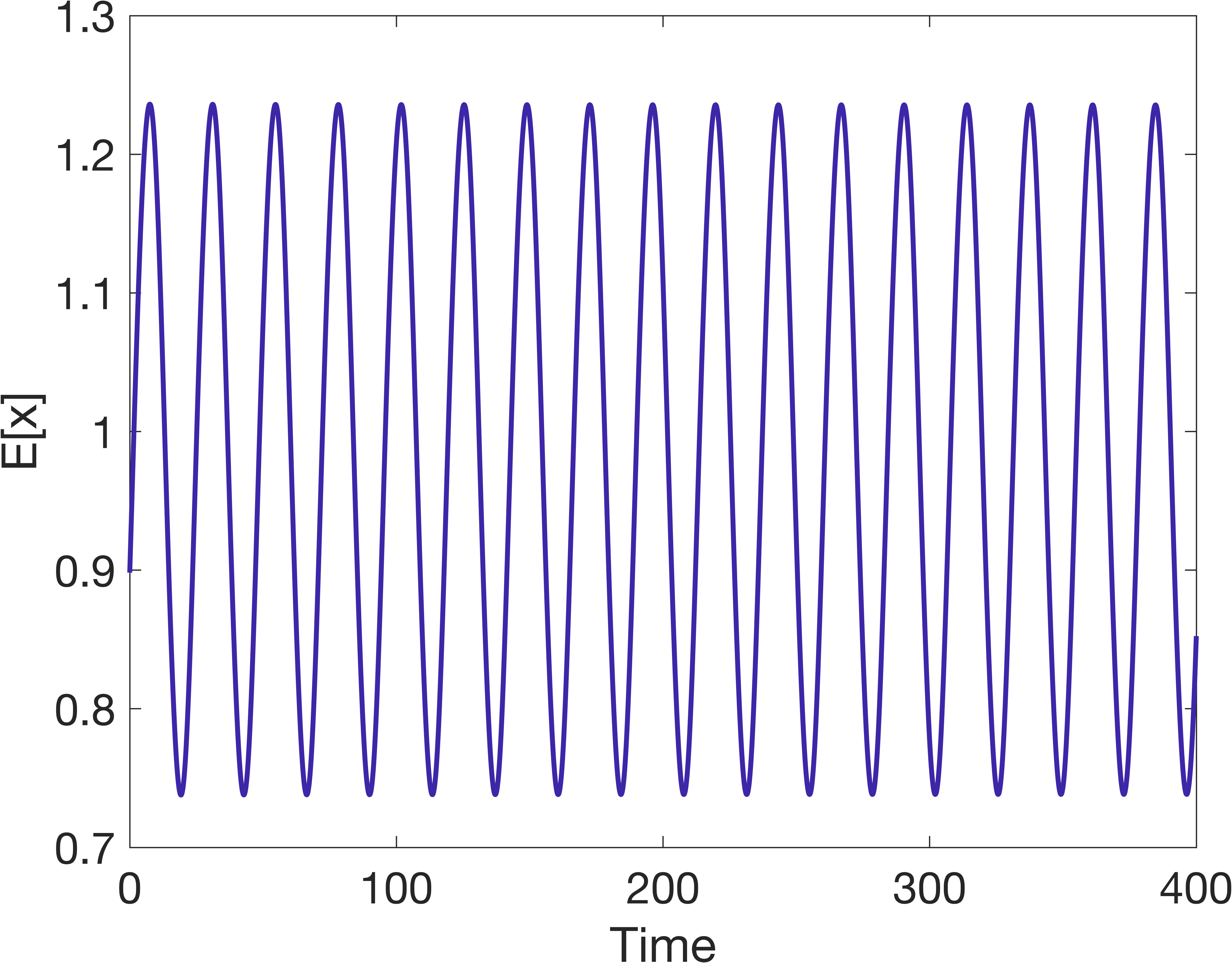} 
  \caption{}
  \label{fig:tsk1}
\end{subfigure}
    \caption{Evolution of the average in $x$ for two limiting scenarios: (a) no coupling with $K=0$ and (b) perfect coupling with $K=1$. $E[x]$ is computed from the solution to the mean-field equation (\ref{eq:mainPDE}).}
    \label{fig:Ex-K}
\end{figure}

\begin{figure}[htbp]
\centering
\begin{subfigure}{\textwidth}
  \centering
  \includegraphics[width=0.7\linewidth]{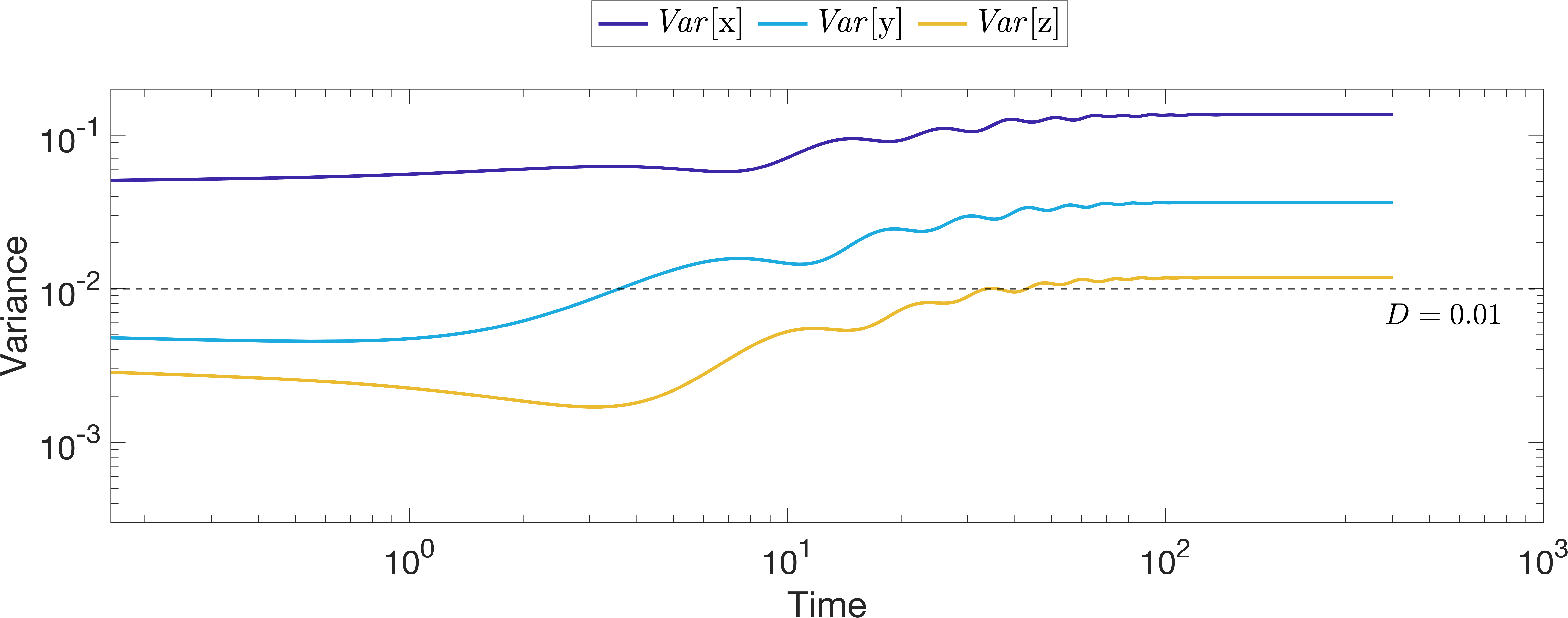} 
  \caption{}
  \label{fig:vark0}
\end{subfigure}
\begin{subfigure}{\textwidth}
  \centering
  \includegraphics[width=0.7\linewidth]{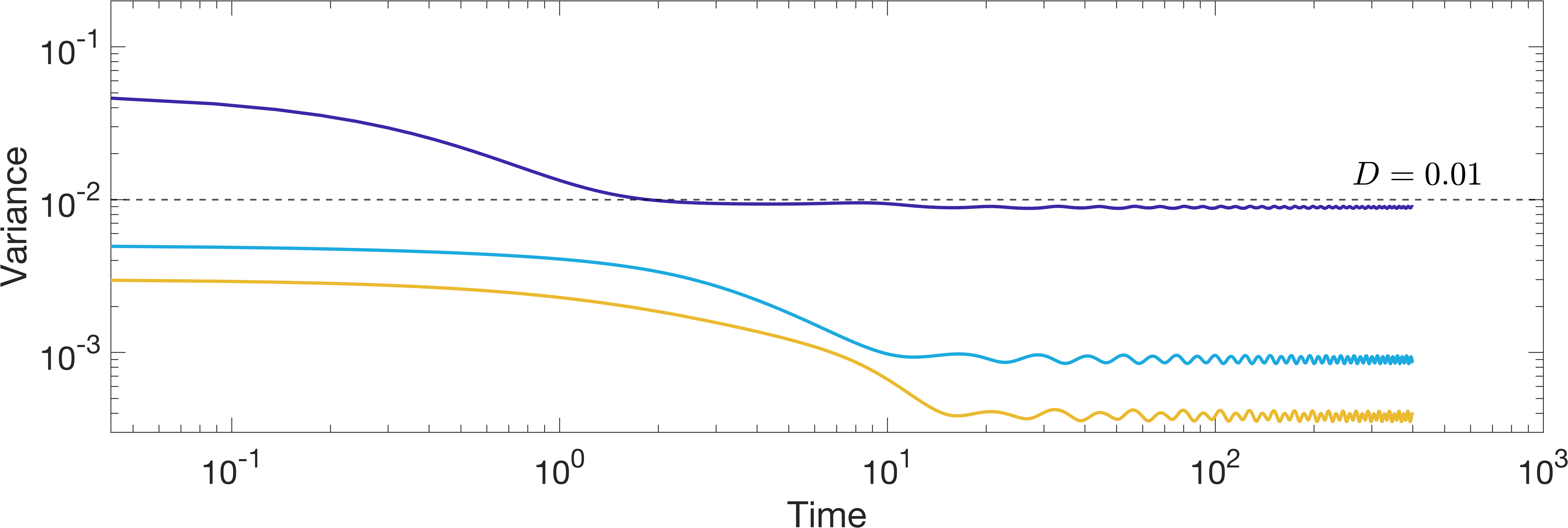}
  \caption{}
  \label{fig:vark1}
\end{subfigure}
\caption{Time evolution of the empirical variance in two limiting scenarios: (a) no coupling with $K=0$ and (b) perfect coupling with $K=1$. $D=0.01$ in all simulations.}
\label{fig:varK2}
\end{figure}

Examining the temporal evolution of the empirical variance offers a complementary perspective on the role of coupling. Synchronization can be understood as the dissipation of the empirical variance for the spatial averages of $x$, $y$ and $z$ computed from the solution of (\ref{eq:mainPDE}) as time goes by. In Fig~\ref{fig:Ex-K} and Fig~\ref{fig:varK2} we show two extreme scenarios in this regard: absence of coupling where $K=0$ and perfect coupling with $K=1$. Fig~\ref{fig:Ex-K} gives the time evolution of $E[x]$. The solution to the mean-field equation approaches a steady state when $K=0$ as shown by damped oscillations in $E[x]$, but looses its stability to rapidly settle into a periodic orbit when $K=1$. The other variables $y$ and $z$, which are not shown here, are qualitatively similar with $x$. It should be noted that regardless of initial conditions, the system follows the same trajectories (data not presented). For each scenario, the time evolution of the variance for all three variables is recorded in Fig~\ref{fig:varK2}. Assuming a low level of noise ($D=0.01$), we observe when $K=0$ an asynchronous state represented by an empirical variance of constant order in time and which is greater than the input noise level. However, for perfect coupling $K=1$, the empirical variance dissipates to a minimum equal to the magnitude of the external noise, and the global output is rhythmic (see Fig~\ref{fig:Ex-K}B and Fig~\ref{fig:varK2}B). In the ideal case of perfect coupling, the variance in $y$ and $z$ decreases to about zero, whereas the variance in $x$ decreases to about $0.01$ (the level of input noise $D$). 

\begin{figure}[htbp]
\begin{subfigure}{.32\textwidth}
  \centering
  \includegraphics[width=\linewidth]{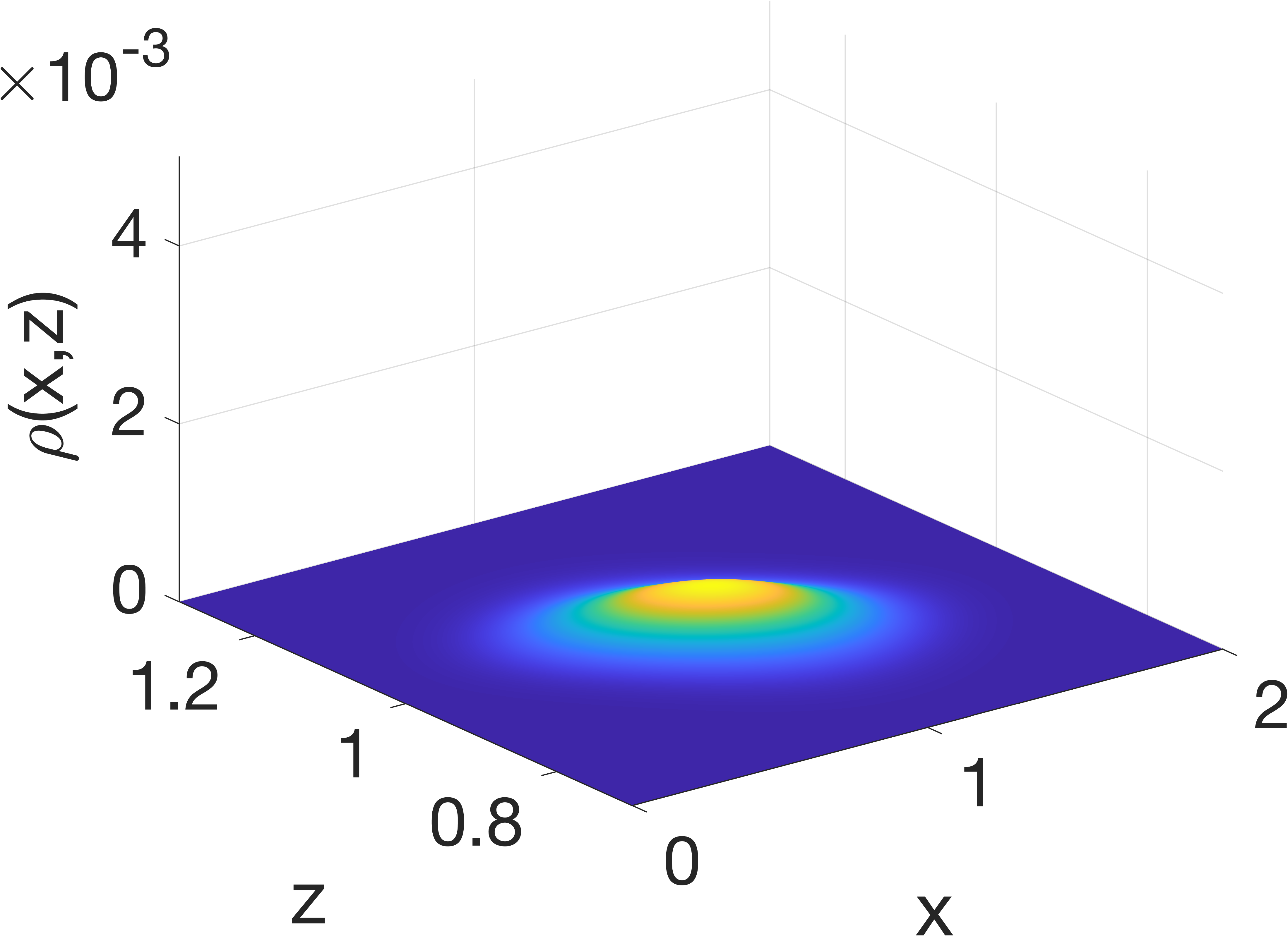} 
  \caption{}
  \label{fig:jointPk01}
\end{subfigure}
\hspace{0.1cm}
\begin{subfigure}{.32\textwidth}
  \centering
  \includegraphics[width=\linewidth]{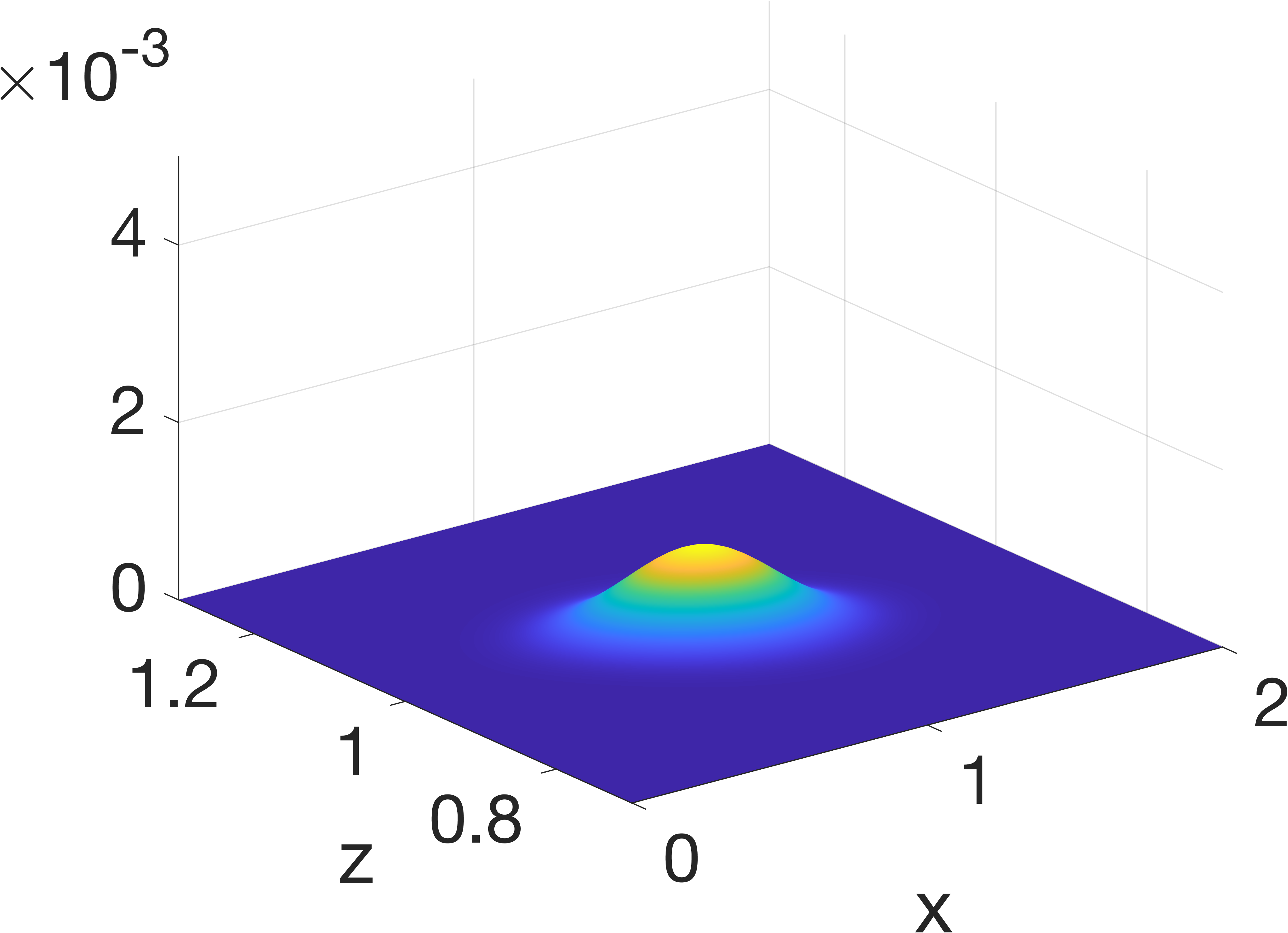} 
  \caption{}
  \label{fig:jointPk02}
\end{subfigure}
\begin{subfigure}{.32\textwidth}
  \centering
  \includegraphics[width=\linewidth]{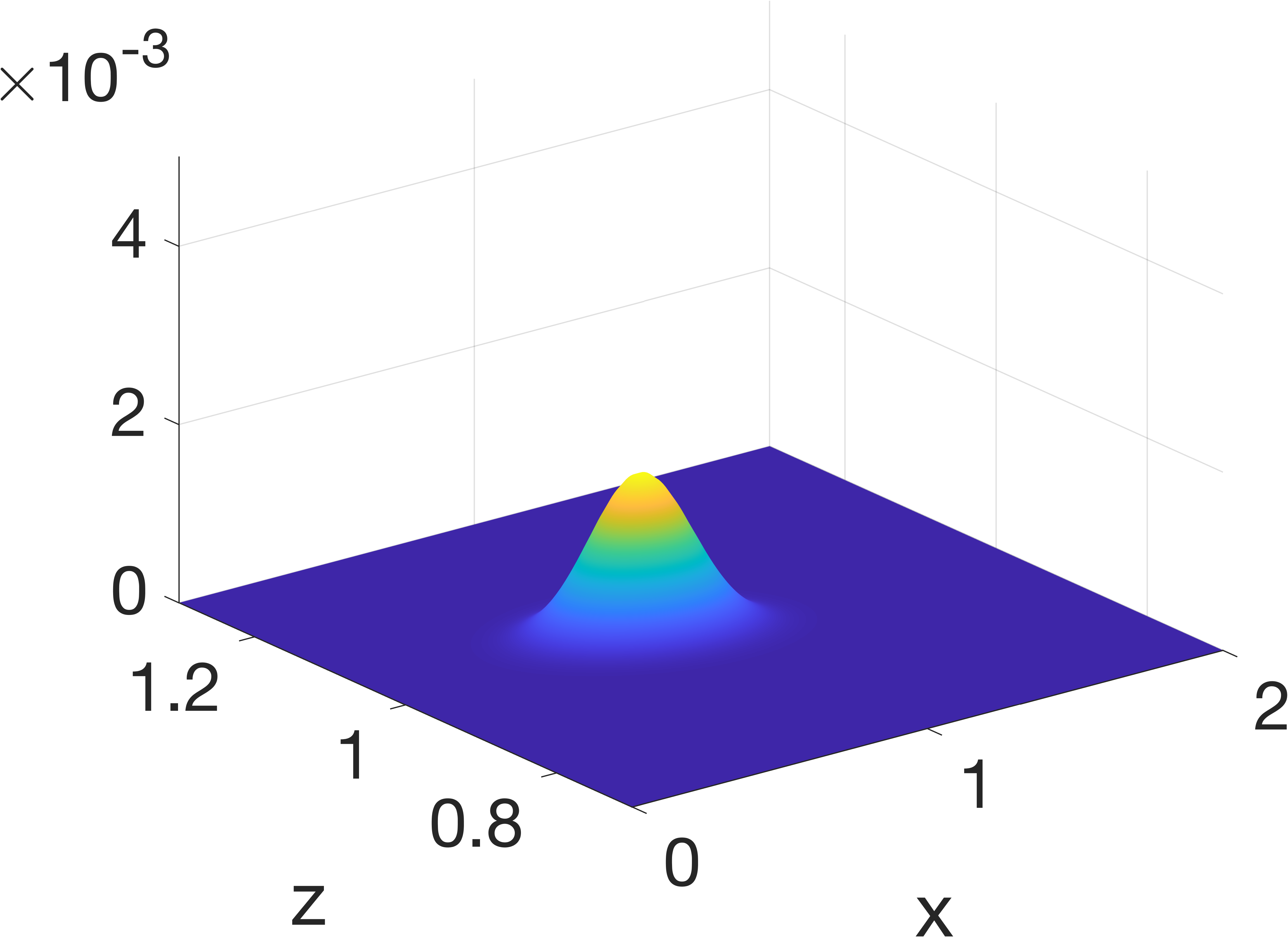} 
  \caption{}
  \label{fig:jointPk04}
\end{subfigure}
\hspace{0.1cm}
\begin{subfigure}{.32\textwidth}
  \centering
  \includegraphics[width=\linewidth]{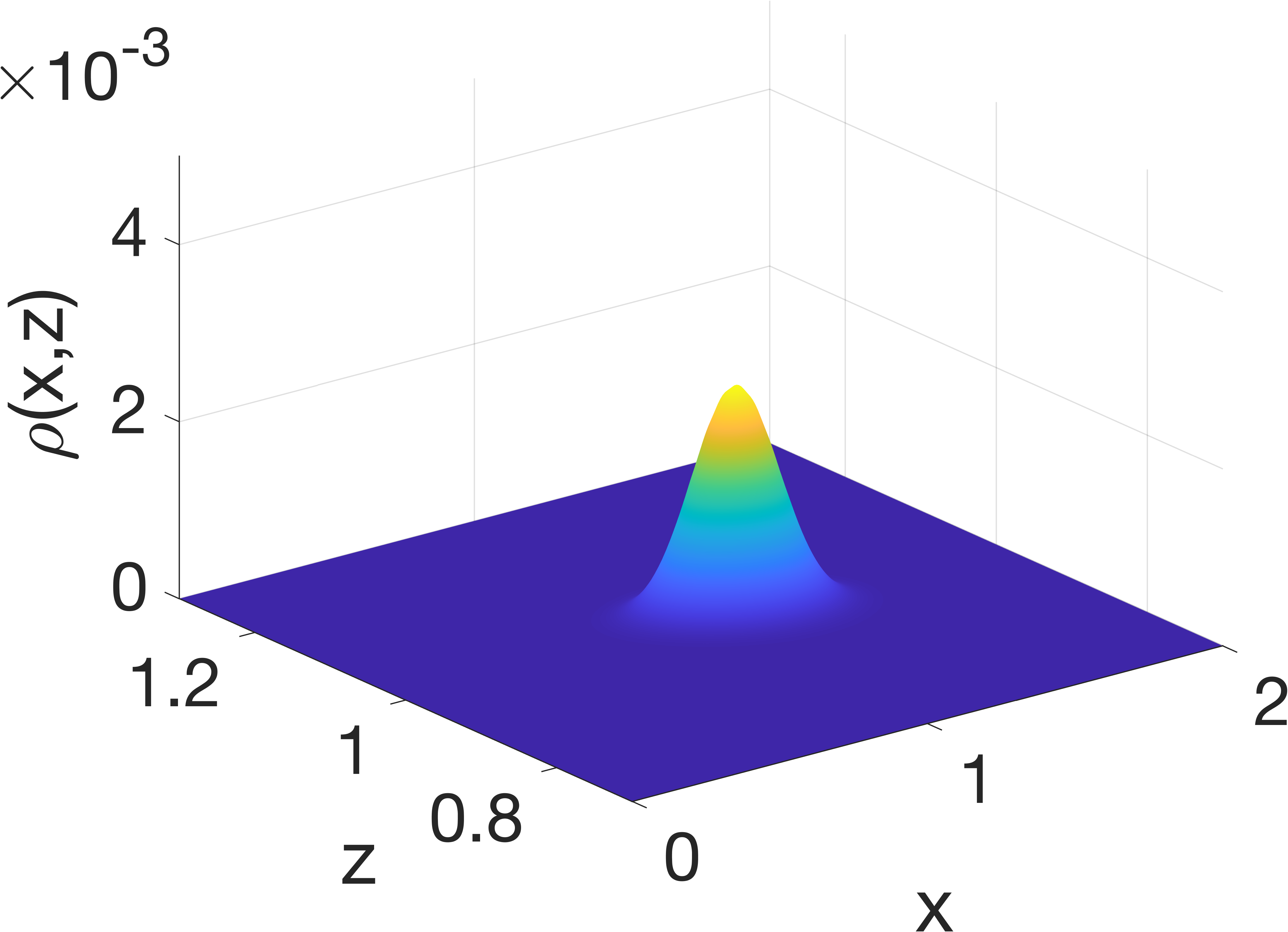} 
  \caption{}
  \label{fig:jointPk06}
\end{subfigure}
\begin{subfigure}{.32\textwidth}
  \centering
  \includegraphics[width=\linewidth]{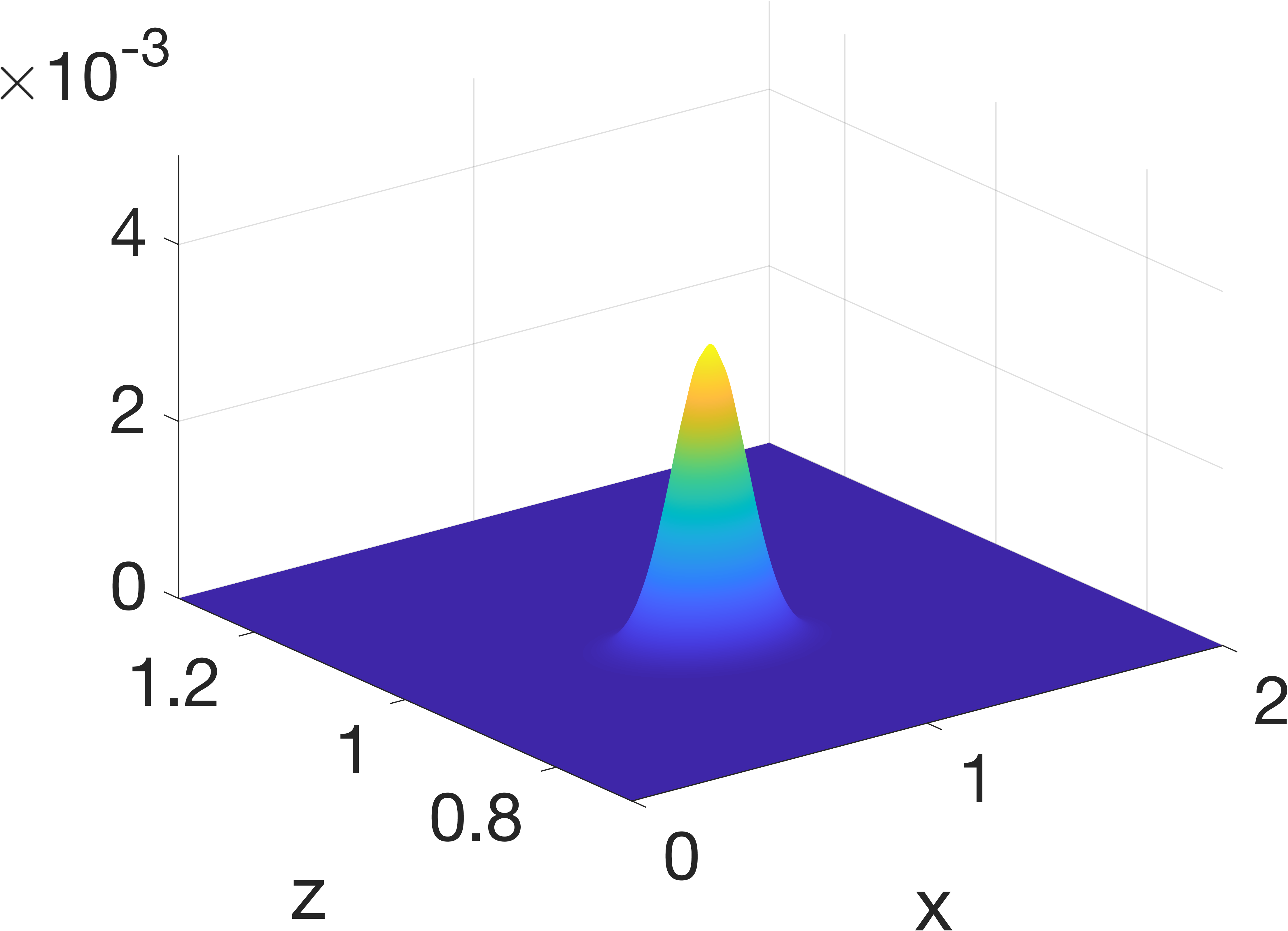} 
  \caption{}
  \label{fig:jointPk08}
\end{subfigure}
\hspace{0.1cm}
\begin{subfigure}{.32\textwidth}
  \centering
  \includegraphics[width=\linewidth]{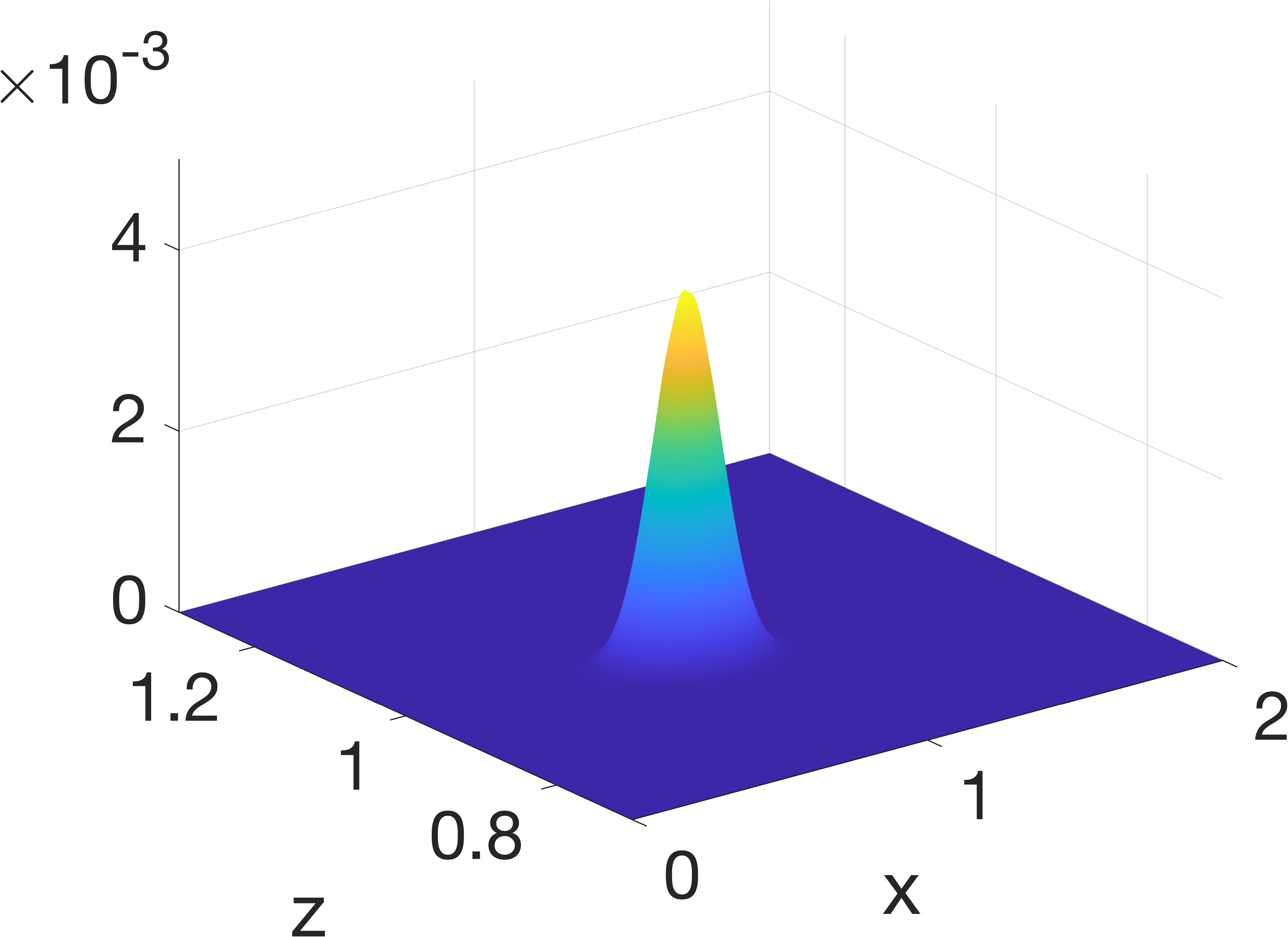} 
  \caption{}
  \label{fig:jointPk1}
\end{subfigure}
\caption{Joint probability distribution between $x$ and $z$ at $t=600$. (A) K=0.1, (B) K=0.2, (C) K=0.4, (D) K=0.6, (E) K=0.8 and (F) K=1. Noise level $D=0.01$.}
\label{fig:jointPk}
\end{figure}

Our numerical experiment suggests the following about the qualitative behavior of the coupled SCN network in terms of $D$: for $D>0$ the neuron trajectories are enclosed in a band whose width rises with $D$, whereas a noise-free network will ultimately reach perfect synchronisation where the empirical variance decreases to zero. The direct consequence of these observations is that the threshold $K_H$ is itself a function of $D$. To see this, consider Fig~\ref{fig:jointPk} which shows the joint probability distribution between $x$ and $z$ at $t=600h$, $\rho(t=600,x,z)$, for different values of $K$. We choose a long integration time in order to not be influenced by the initial conditions. 
As the coupling parameter increases, variance decreases and the distribution $\rho(t=600,x,z)$ tends to concentrate on a delta function in the $z$ dimension and to be distributed only along the spatial dimension $x$ due to noise. 
Overall, studying the effects of coupling on the solution to the mean-field limit gave the following numerical predictions: increasing coupling strength can lead to period locking, variance dissipation, and larger amplitudes due to resonance effects.

\subsection{Effect of noise on bifurcation boundaries} \label{sec:noise}
Using our stochastic mean-field model, we examine how the robustness of the SCN clock is affected by noise. This refers to how the noise intensity in the system affects the distance to a bifurcation point. We proceed by varying the parameter $\alpha $ in (\ref{eq:mainPDE}) for different noise levels, and we look for synchronized activity within the network. Robustness is used here to denote the persistence of a certain type of dynamic behavior over a significant range of parameter values.
The term ``robustness'' refers to the persistence of a specific dynamic behavior over a wide range of parameter values.

\begin{figure}[htbp]
\centering
\begin{subfigure}{.48\textwidth}
  \centering
  \includegraphics[width=\linewidth]{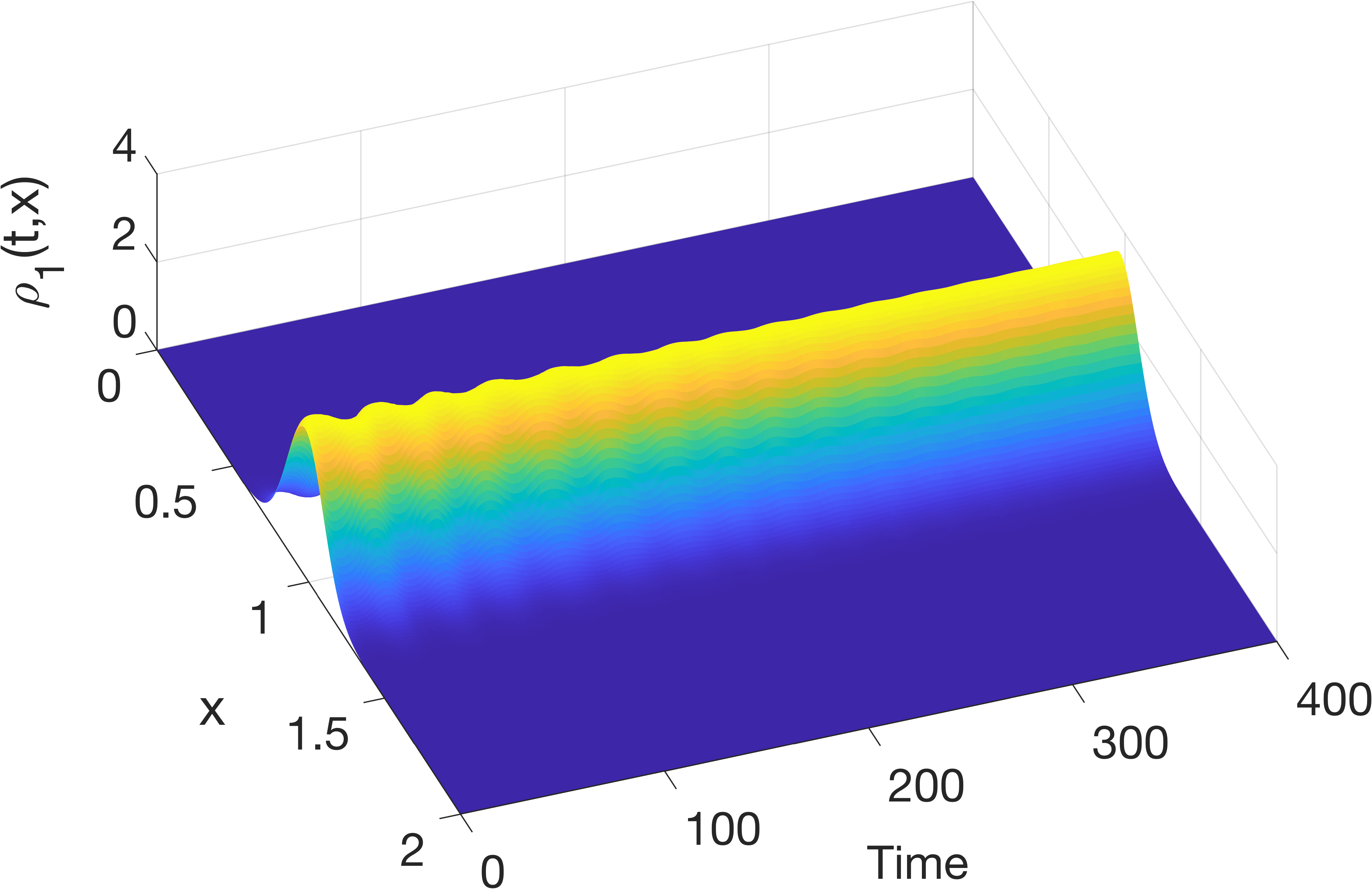}
  \caption{steady state regime}
  \label{fig:low15x}
\end{subfigure}
\hspace{0.1cm}
\begin{subfigure}{.48\textwidth}
  \centering
  \includegraphics[width=\linewidth]{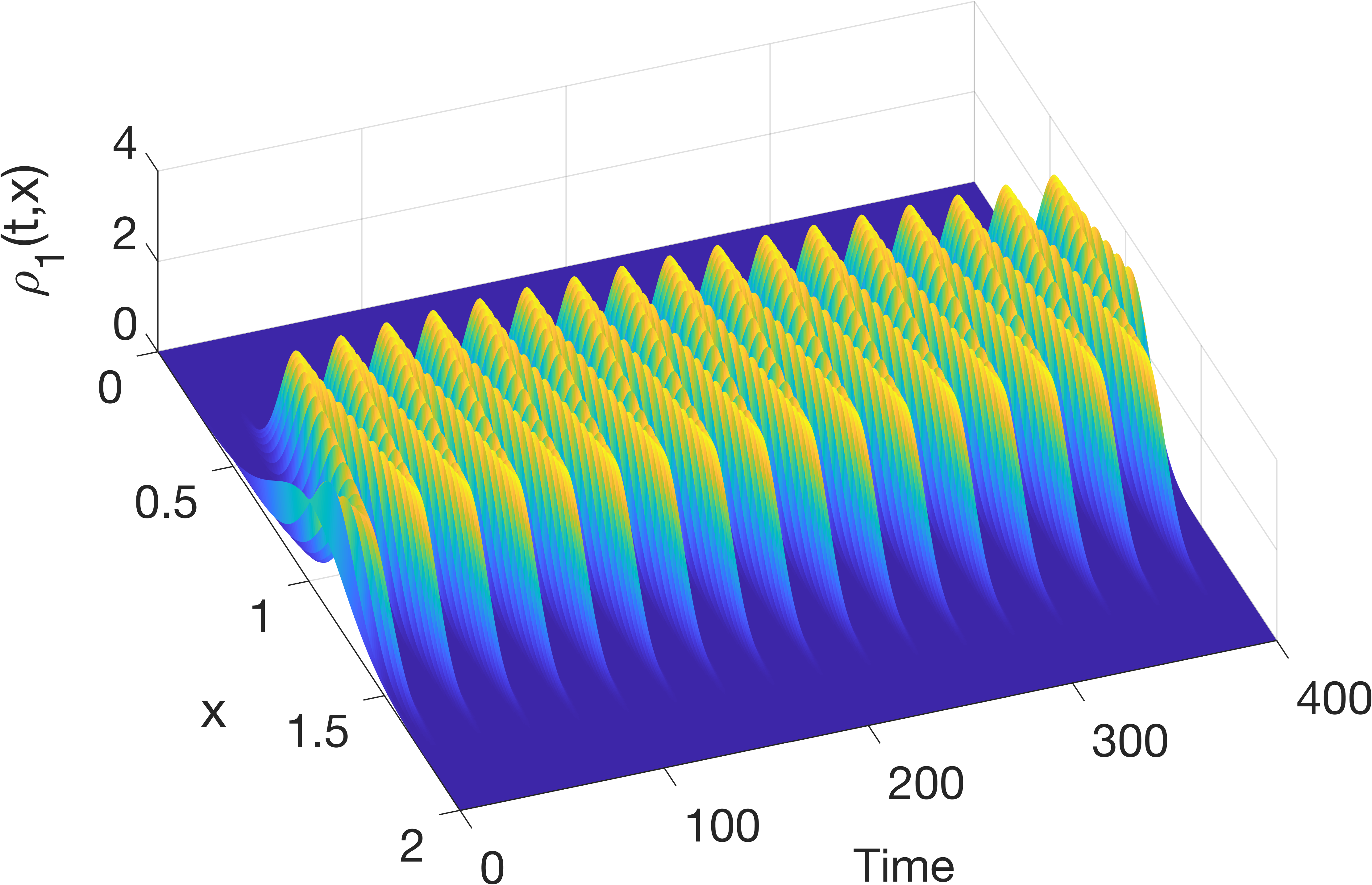}
  \caption{oscillatory regime}
  \label{fig:low20x}
\end{subfigure}
\caption{Evolution of the marginal density of $x$ in the presence of noise. (a) steady state regime with $\alpha=1.5$, (b) oscillatory regime with $\alpha = 2$. Other parameters: $n=20$, $K=0.6$, $D=0.01$.}
\label{fig:evol-lownoise}
\end{figure}

Consider $D=0.01$ as an example of low noise setting. As shown in Fig~\ref{fig:evol-lownoise}, there is in the low noise setting a critical value $\alpha_H$ at which the system's stability appears to change from a stable stationary distribution to an oscillatory solution. This result shows that the development and maintenance of a global rhythmic output requires the synchronization of single-cell rhythms \cite{gonze2006circadian}. The existence of such invariant distributions characterizes a state of incoherence within the SCN network (see Fig~\ref{fig:low15x}). Noting that $\alpha = k_1k_3k_5/k_2^3K_i$, Fig~\ref{fig:evol-lownoise} suggests that the circadian system is more favorable to lower degradation rates in the presence of noise.

\begin{figure}[htbp]
\centering
\begin{subfigure}{.48\textwidth}
  \centering
    \includegraphics[width=\linewidth]{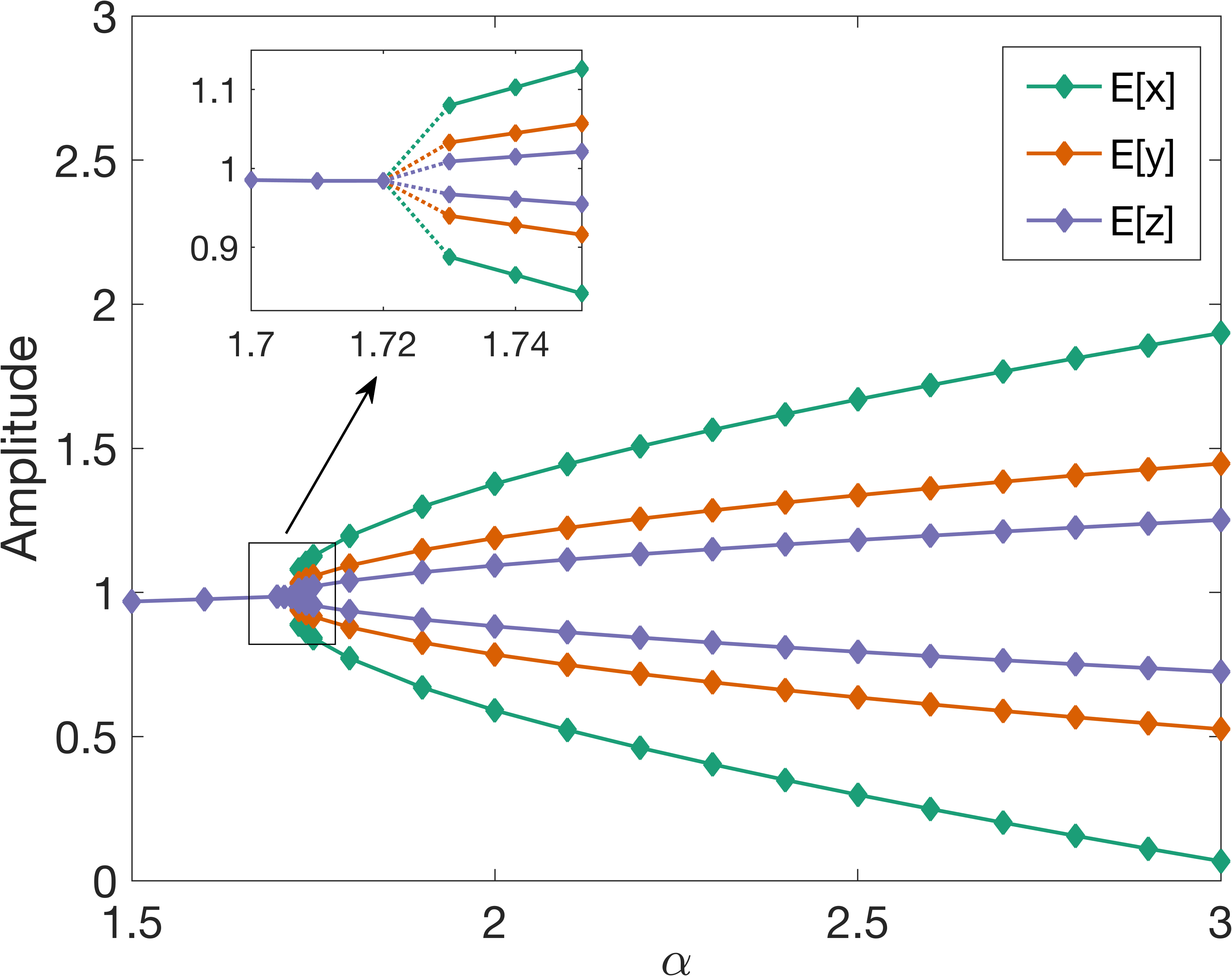} 
    \caption{}
    \label{fig:hopf_low_inset}
\end{subfigure}
\hspace{0.1cm}
\begin{subfigure}{.48\textwidth}
  \centering
    \includegraphics[width=\linewidth]{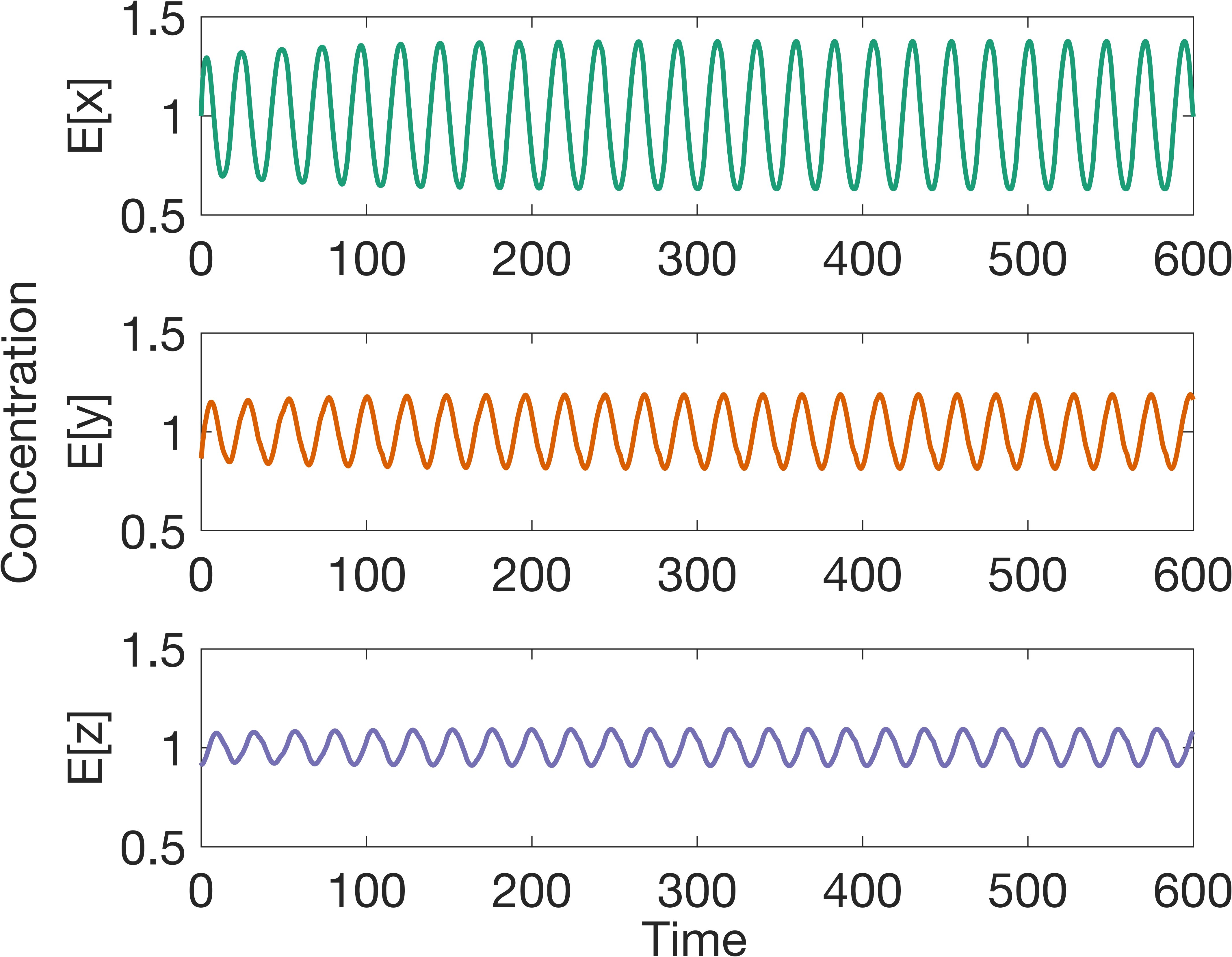} 
    \caption{}
    \label{fig:orbit-low20}
\end{subfigure}
\caption{(a) Bifurcation diagram associated with the parameter $\alpha$. Details of the diagram are the same as those of Fig~\ref{fig:bifK2}. (b) Example of a periodic orbit when $\alpha=2$. Other parameters: $n=20$, $K=0.6$, $D=0.01$.}
\label{fig:bif-lownoise}
\end{figure}

To formally investigate the bifurcation in Fig~\ref{fig:evol-lownoise}, we use the same parameters and initial conditions as in Fig~\ref{fig:evol-lownoise}, with the exception of the parameter $\alpha$ which now varies from 1.5 to 3. Results are displayed in Fig~\ref{fig:hopf_low_inset}. When $\alpha < 1.73$, the network is not synchronised and evolves towards a steady state. Above the critical value $\alpha_H \approx 1.73$, the network is synchronised and the solution to the mean-field equation (\ref{eq:mainPDE}) is periodic in time. As characteristic of a local Hopf bifurcation, the cycle that is born is nearly elliptical with a small amplitude; see Fig~\ref{fig:orbit-low20}. 

\begin{figure}[htbp]
\begin{subfigure}{.48\textwidth}
  \centering
  \includegraphics[width=\linewidth]{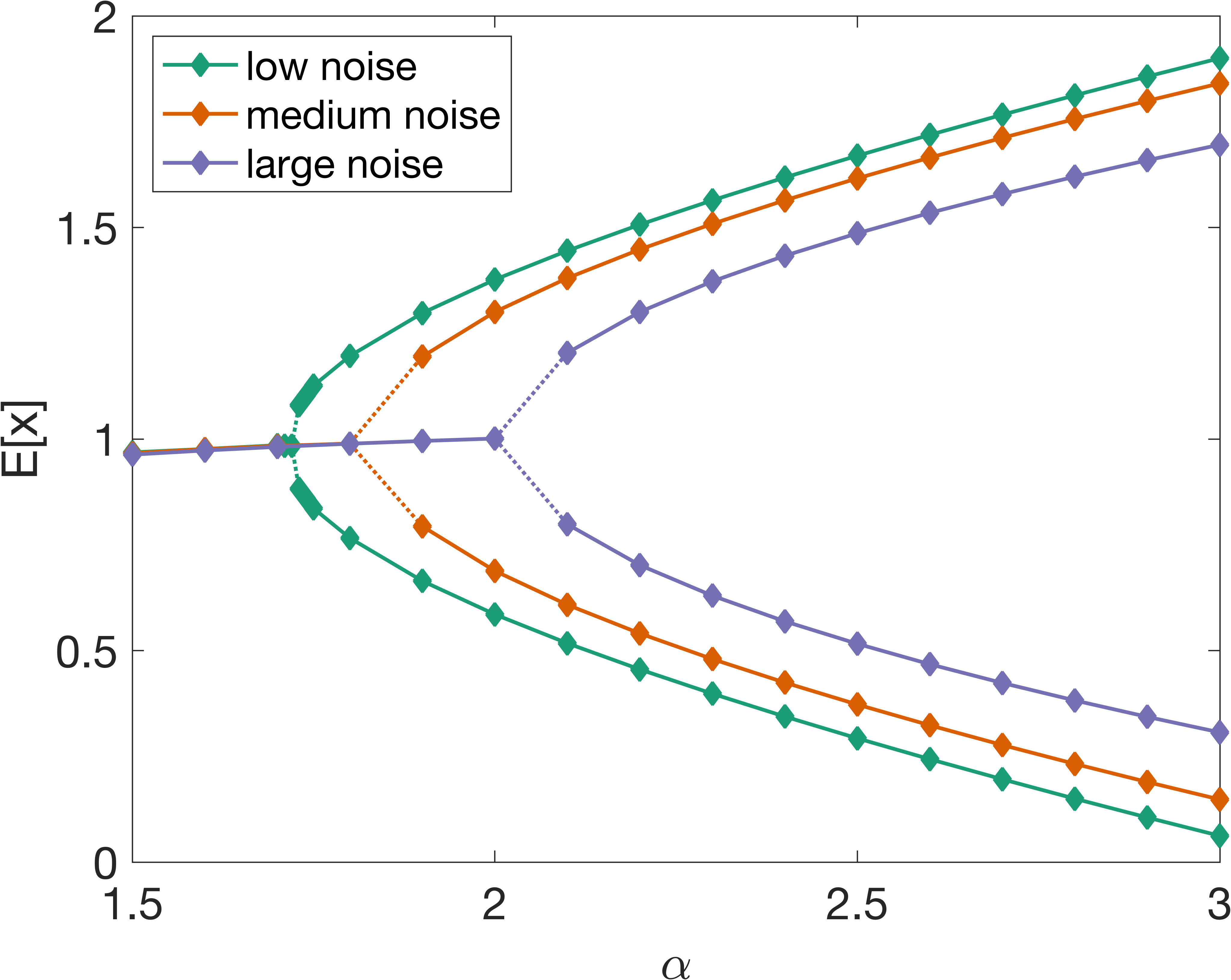} 
  \caption{}
  \label{fig:EXbif}
\end{subfigure}
\hspace{0.1cm}
\begin{subfigure}{.48\textwidth}
  \centering
  \includegraphics[width=\linewidth]{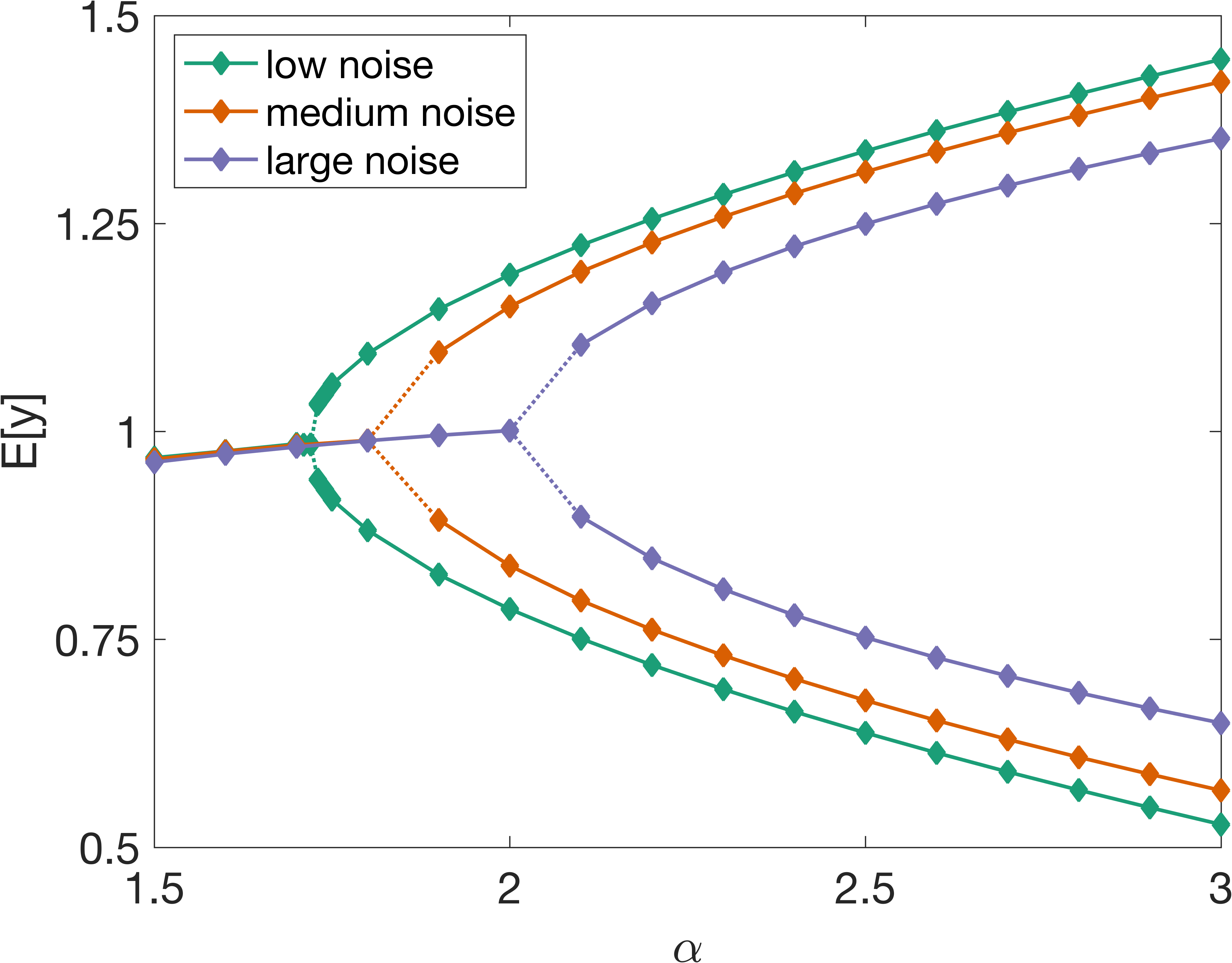} 
  \caption{}
  \label{fig:EYbif}
\end{subfigure}
\par
\begin{subfigure}{.48\textwidth}
  \centering
  \includegraphics[width=\linewidth]{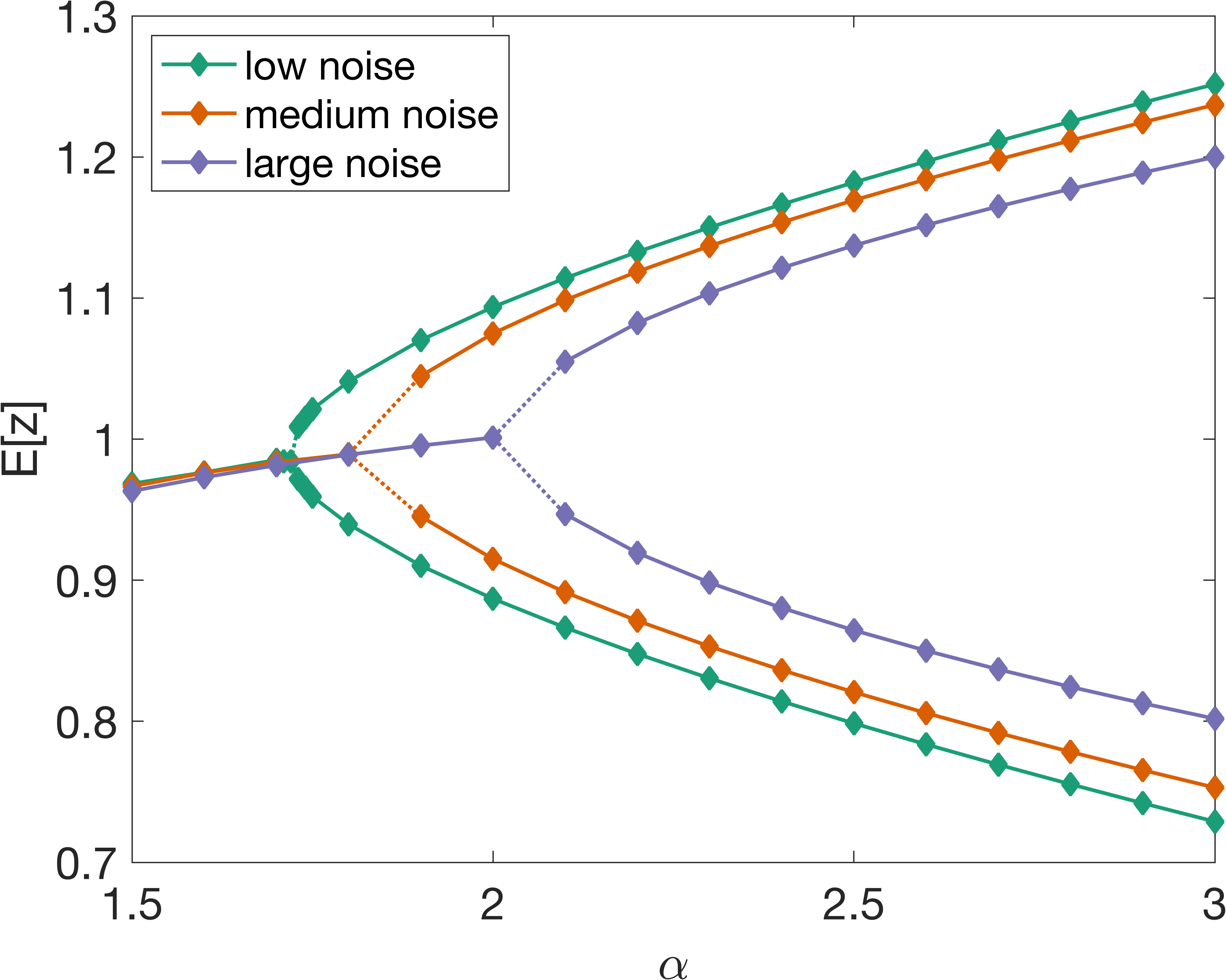} 
  \caption{}
  \label{fig:EZbif}
\end{subfigure}
\hspace{0.1cm}
\begin{subfigure}{.48\textwidth}
  \centering
  \includegraphics[width=\linewidth]{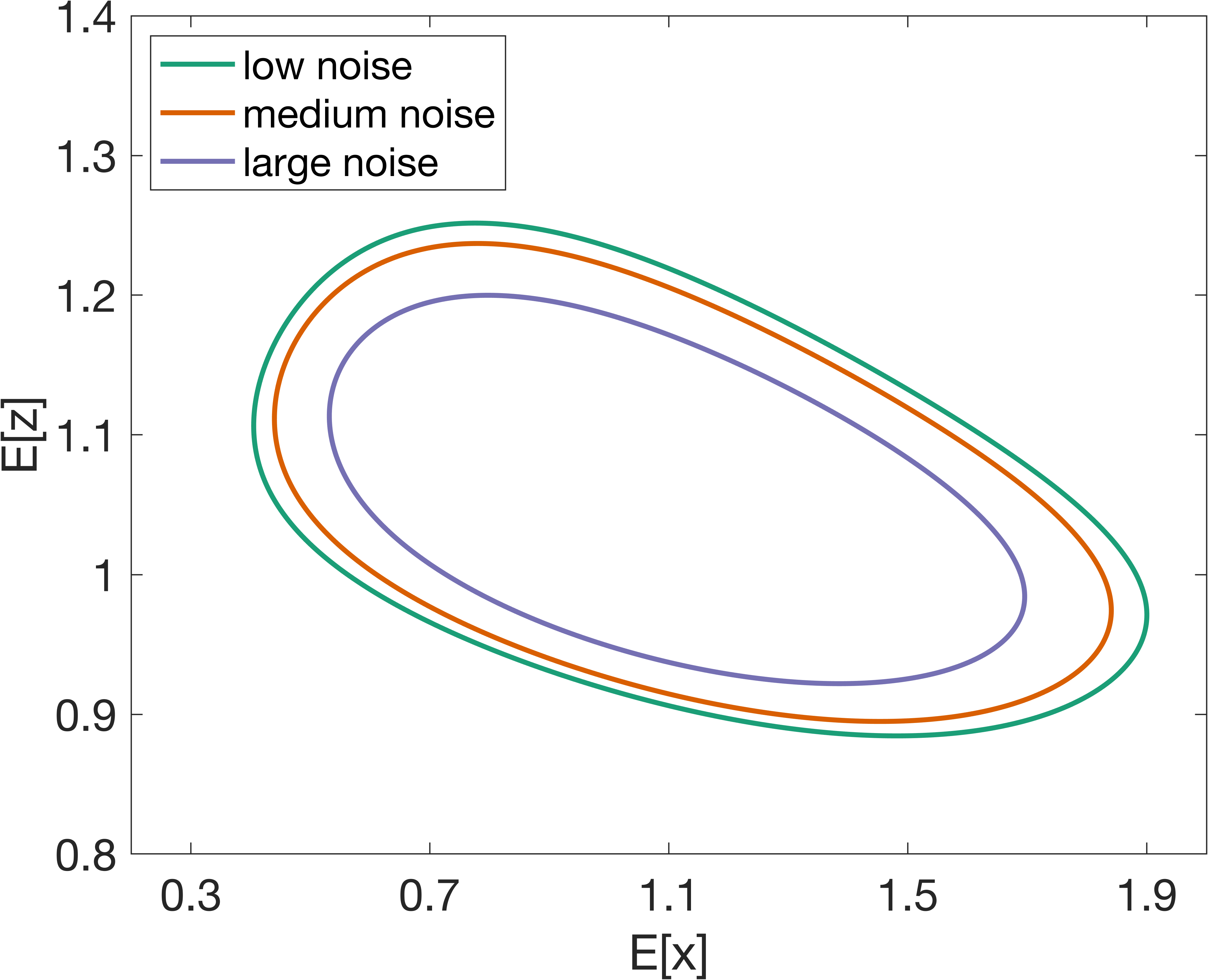} 
  \caption{}
  \label{fig:cycle}
\end{subfigure}
\caption{Bifurcation diagrams associated with the parameter $\alpha$ for the spatial averages (a) $E[x]$, (b) $E[y]$ and (c) $E[z]$. (d) Stable limit cycles in the $E[x]$--$E[z]$ plane when $\alpha = 3$. We model low noise ($D = 0.01$), medium noise($D = 0.025$) and high noise ($D = 0.05$). Dotted lines represent intervals containing exact bifurcation values $\alpha_H$. Details of the figures are the same as those of Fig~\ref{fig:bifK2}.}
\label{fig:noise-bif}
\end{figure}

We investigate further how increasing the level of noise affects bifurcation boundaries. Fig~\ref{fig:noise-bif} shows a high-low plot for the peaks and troughs of oscillations of the means in $x$, $y$ and $z$ as a function of the parameter $\alpha$ for different noise levels. Since it is difficult to numerically estimate exact bifurcation values, we use dotted lines to represent intervals containing exact bifurcation points, which we call $\alpha_H$. We estimate $\alpha_H \in (1.72, 1.73]$, $(1.8, 1.9]$ or $(2.0,2.1]$ when the noise intensity is low, moderate or high, respectively. As the noise level increases, the value of $\alpha_H$ necessary to obtain sustained circadian oscillations also increases. From Fig~\ref{fig:cycle}, the amplitude of these oscillations gradually decreases as noise intensity increases, and the oscillations disappear through a supercritical Hopf bifurcation, thus resulting in trivial behavior with a single fixed point.

These findings indicate that higher noise levels render synchronisation more difficult to achieve. That is because a new and larger value $\alpha_H$ becomes the \textit{sine qua non} condition for oscillations. Given the biological meaning of $\alpha$ in (\ref{eq:goodwin-nondim}), its value could increase to recover oscillations if: 1) activation rate $k_1$, $k_3$, or $k_5$ increases for $x$, $y$ or $z$, respectively; 2) degradation rates decrease for all three clock components ($k_2, k_4, k_6$ where $k_2=k_4=k_6$); 3) inhibition of $x$ by $z$ is attenuated (i.e., lower $K_i$).

\begin{figure}[htbp]
\begin{subfigure}{.32\textwidth}
  \centering
  \includegraphics[width=\linewidth]{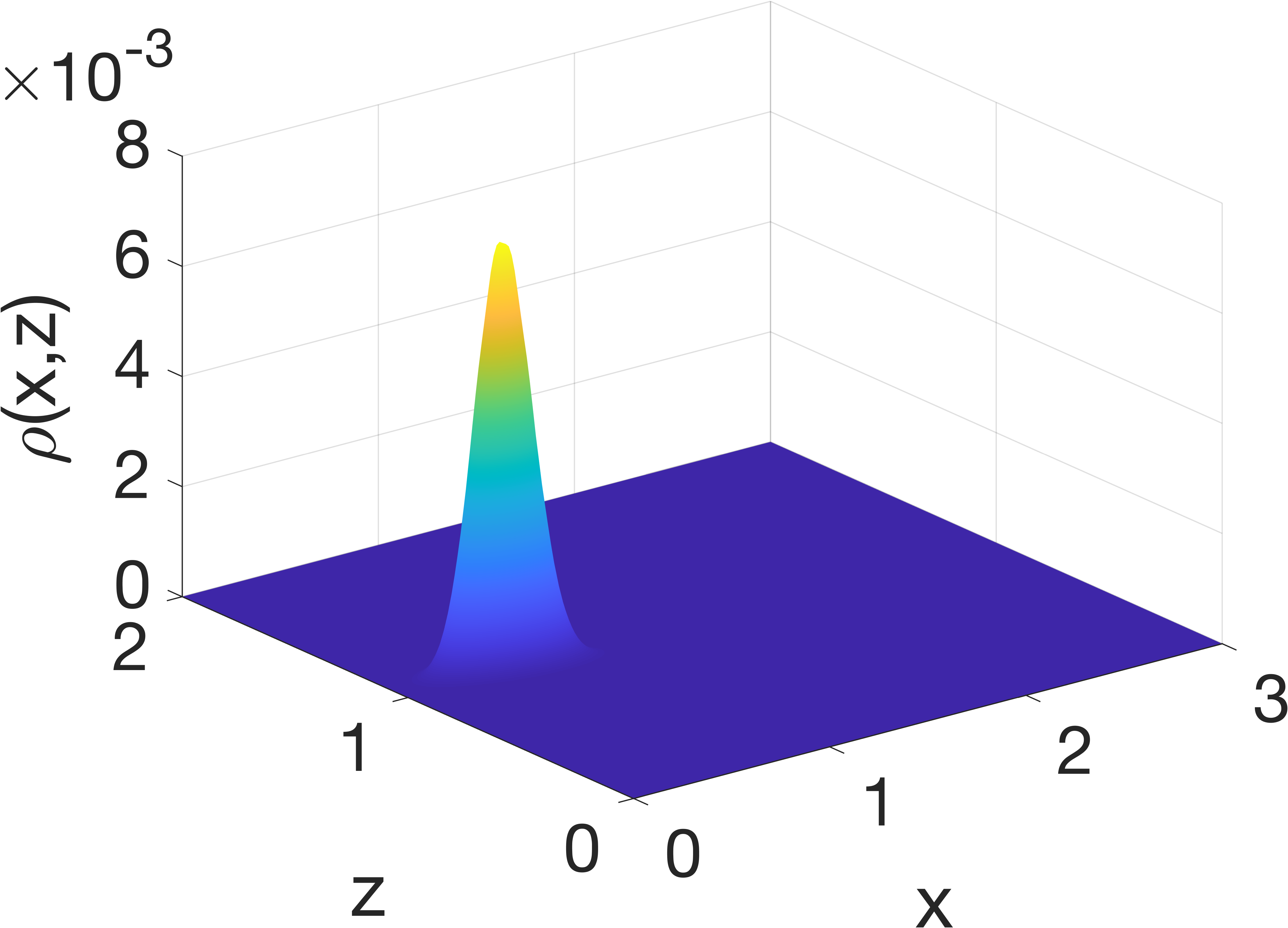} 
  \caption{}
  \label{fig:jointPd01}
\end{subfigure}
\hspace{0.1cm}
\begin{subfigure}{.32\textwidth}
  \centering
  \includegraphics[width=\linewidth]{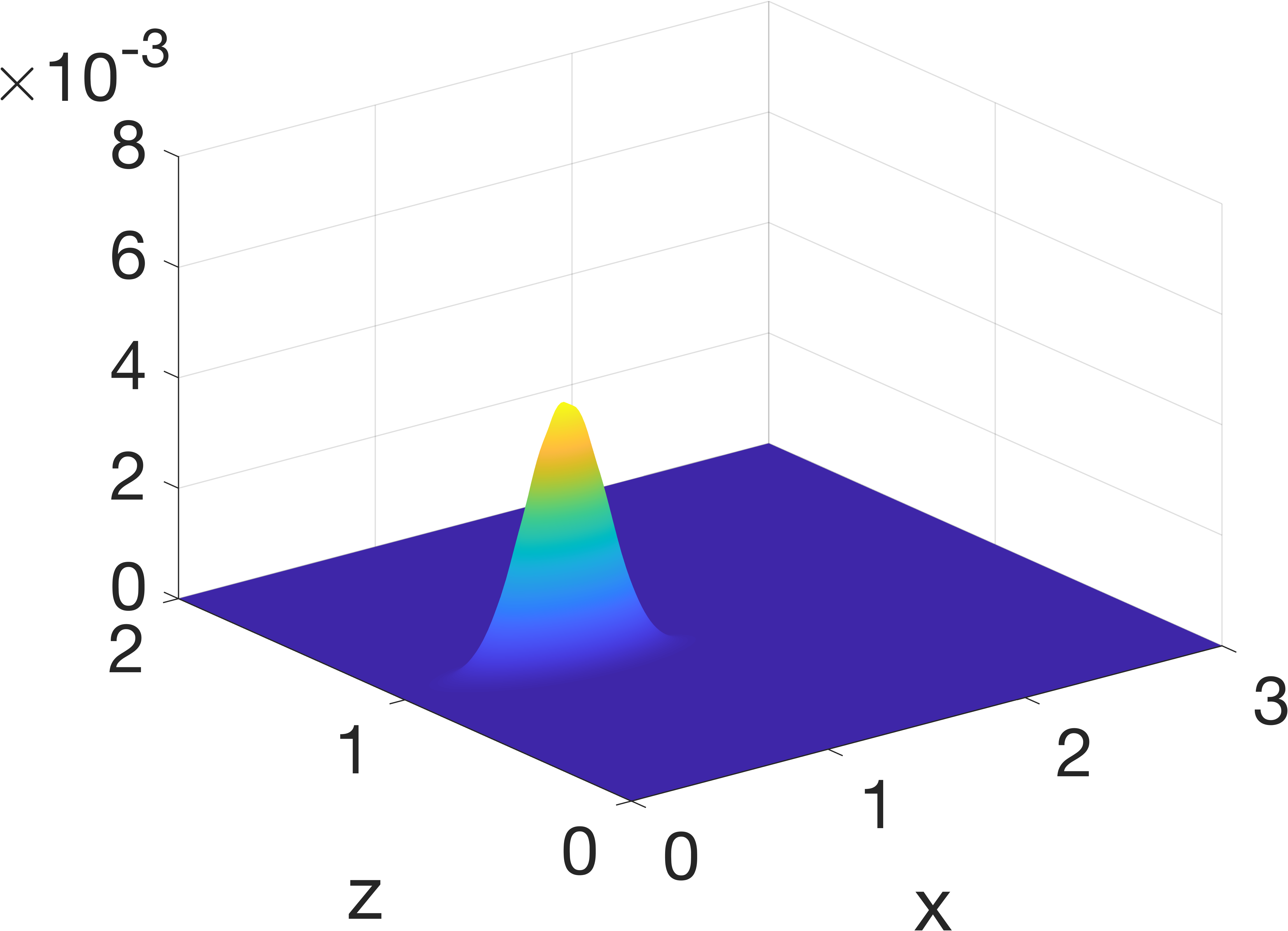} 
  \caption{}
  \label{fig:jointPd02}
\end{subfigure}
\begin{subfigure}{.32\textwidth}
  \centering
  \includegraphics[width=\linewidth]{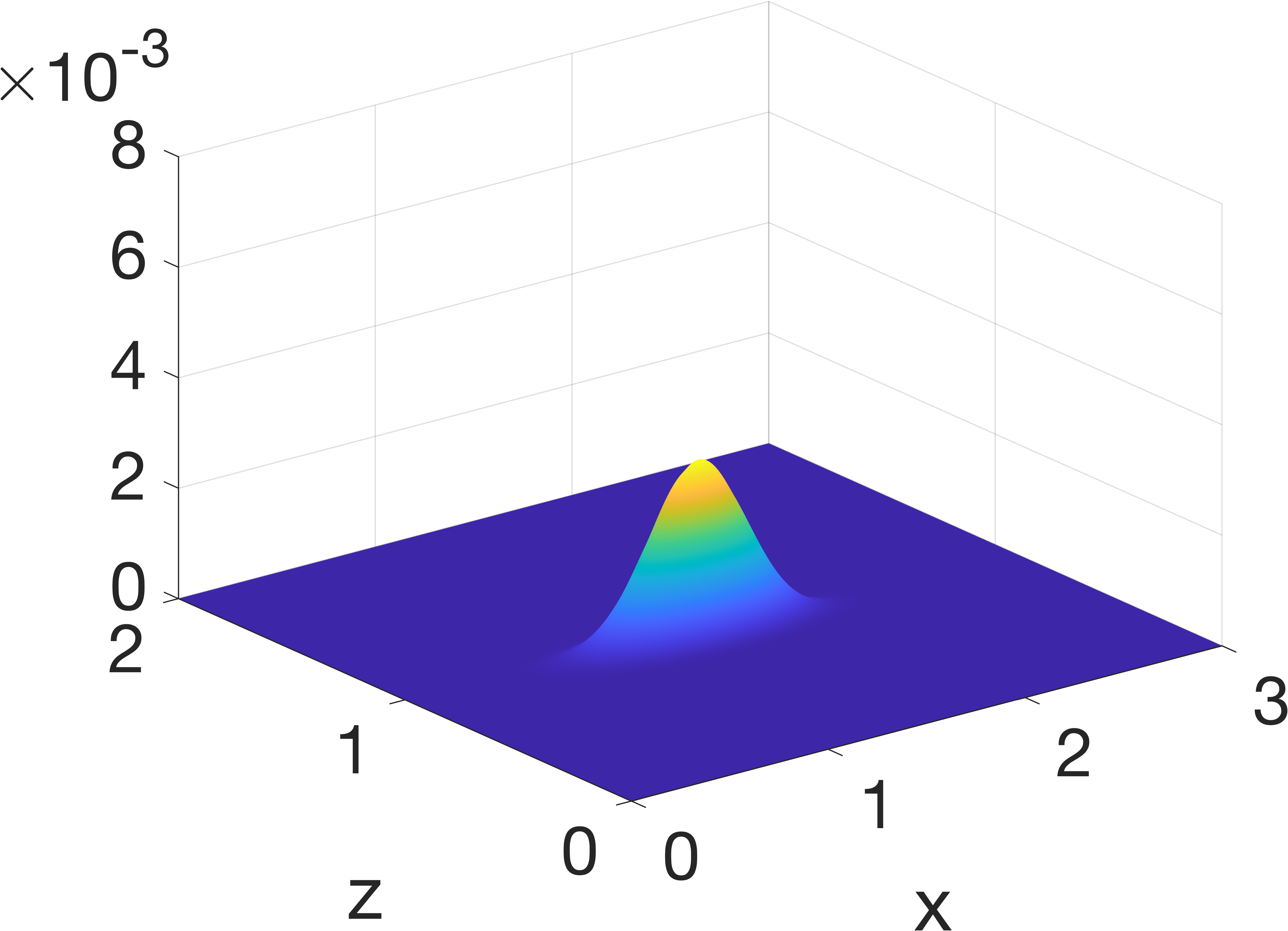} 
  \caption{}
  \label{fig:jointPd04}
\end{subfigure}
\hspace{0.1cm}
\begin{subfigure}{.32\textwidth}
  \centering
  \includegraphics[width=\linewidth]{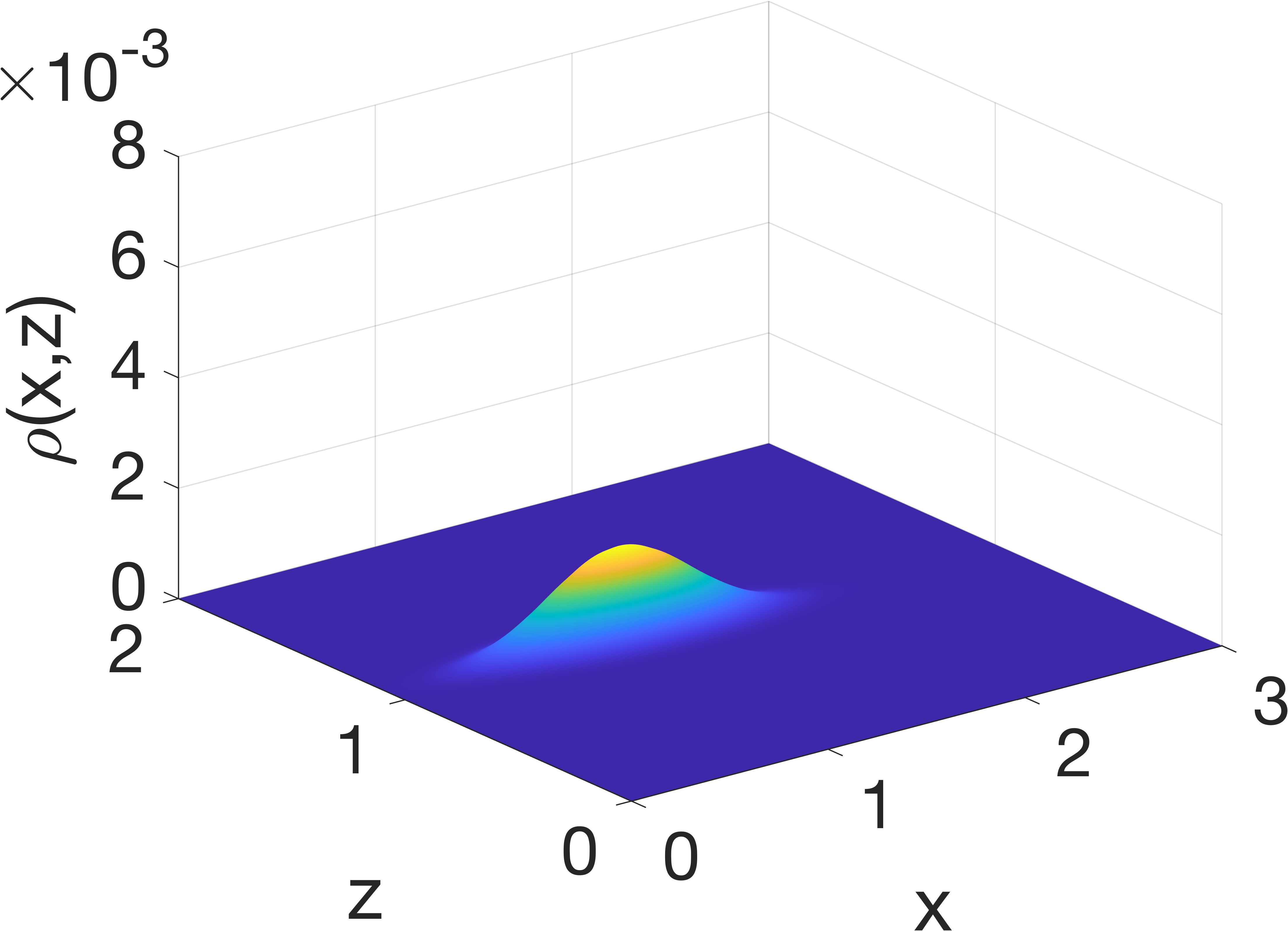} 
  \caption{}
  \label{fig:jointPd08}
\end{subfigure}
\begin{subfigure}{.32\textwidth}
  \centering
  \includegraphics[width=\linewidth]{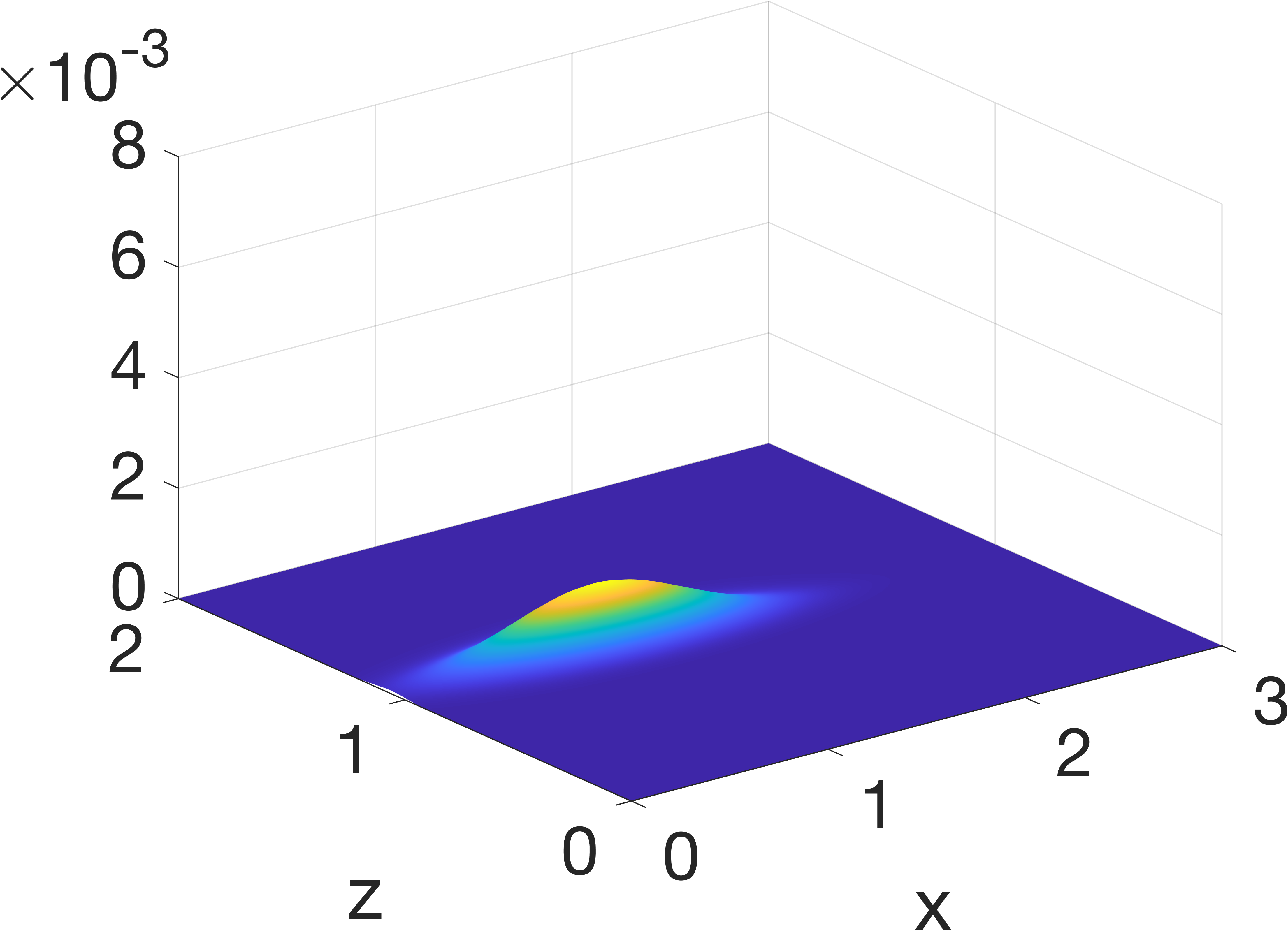} 
  \caption{}
  \label{fig:jointPd12}
\end{subfigure}
\hspace{0.1cm}
\begin{subfigure}{.32\textwidth}
  \centering
  \includegraphics[width=\linewidth]{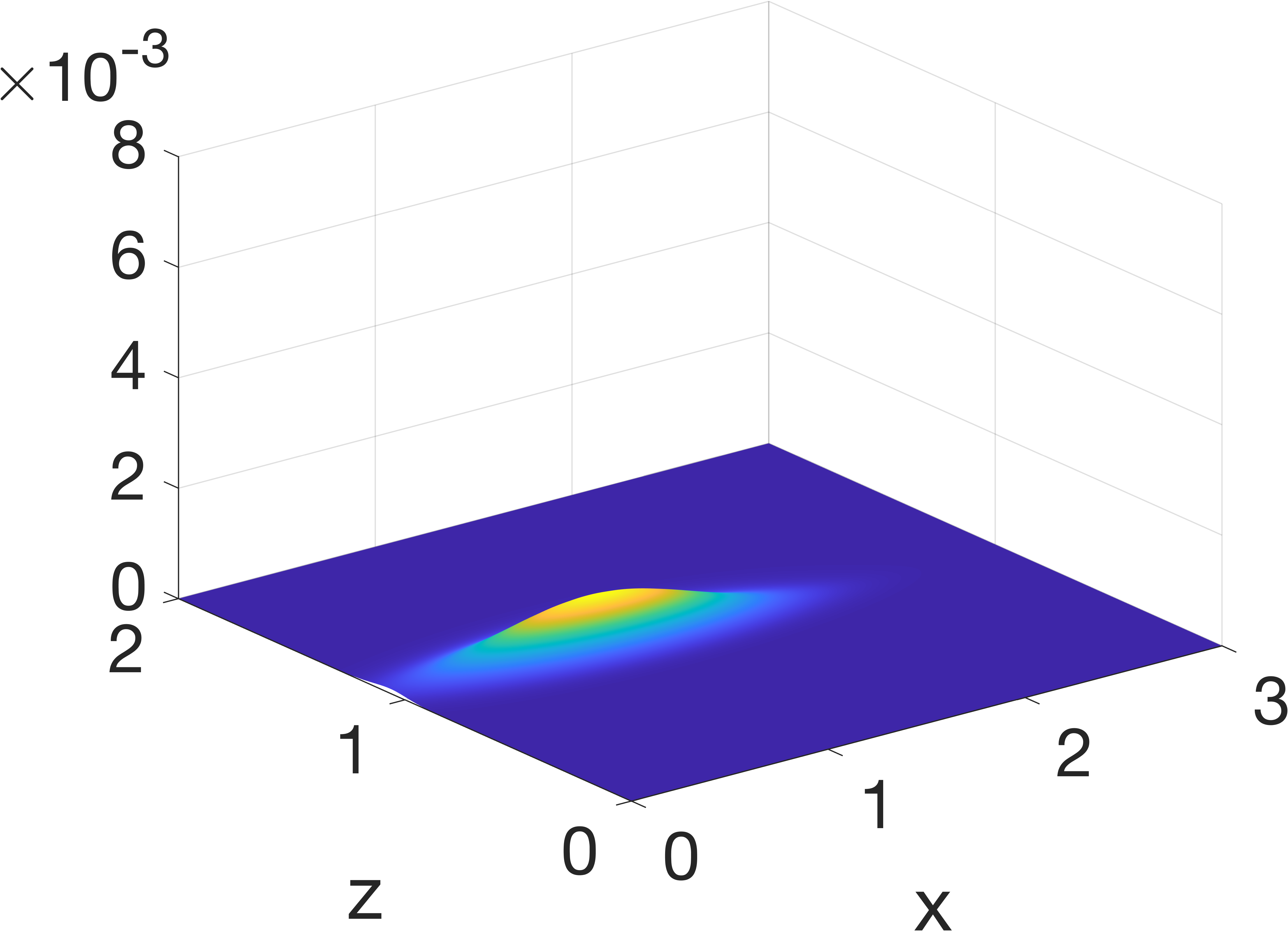} 
  \caption{}
  \label{fig:jointPd15}
\end{subfigure}
\caption{Joint probability distribution between $x$ and $z$ at $t=600$. (a) $D=0.01$, (b) $D=0.02$, (c) $D=0.04$, (d) $D=0.08$, (e) $D=0.12$ and (f) $D=0.15$. Coupling strength $K=0.6$.}
\label{fig:jointPd}
\end{figure}

\begin{figure}[htbp]
\centering
\begin{subfigure}{.48\textwidth}
  \centering
  \includegraphics[width=\linewidth]{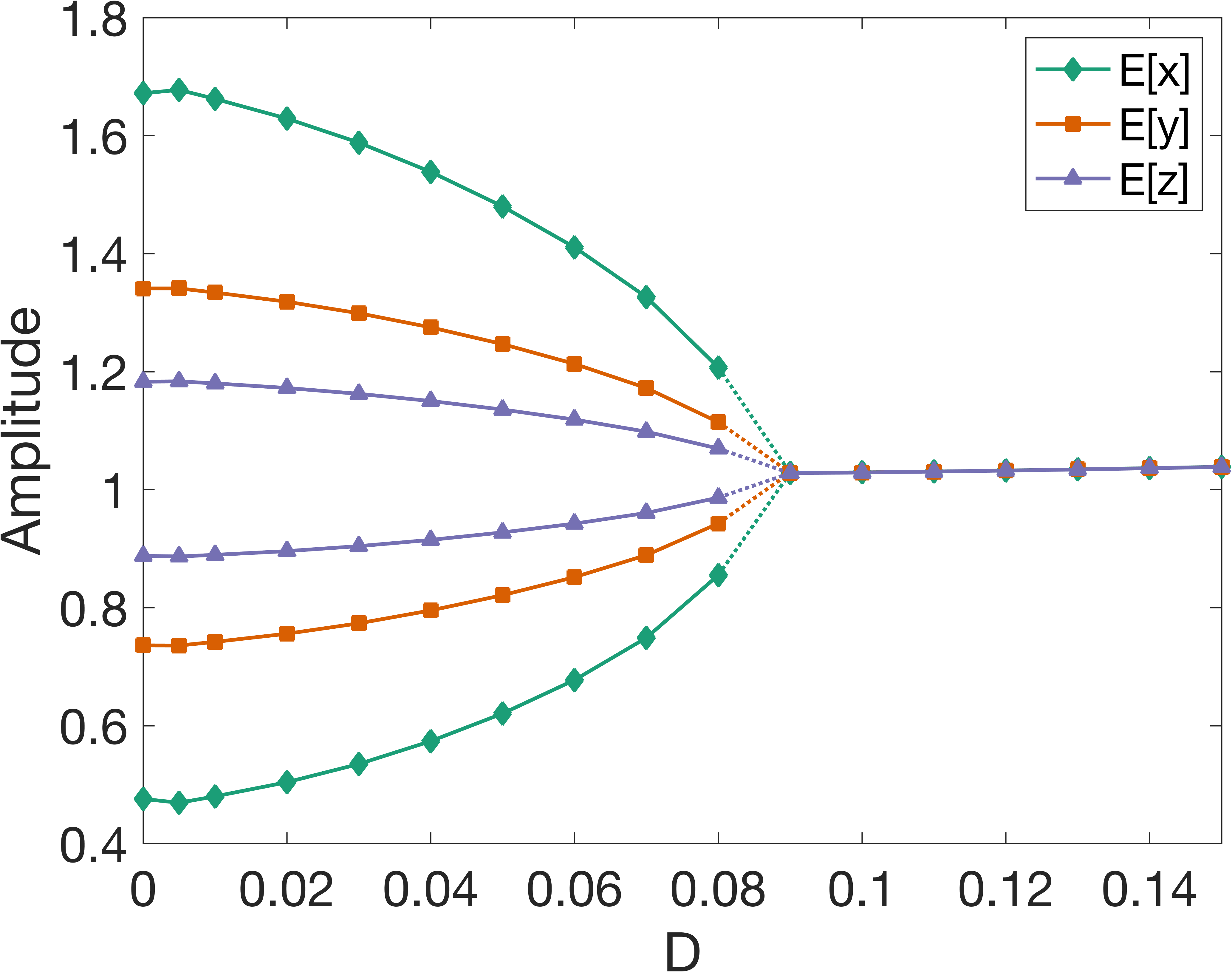}
  \caption{}
  \label{fig:bifD}
\end{subfigure}
\hspace{0.1cm}
\begin{subfigure}{.48\textwidth}
  \centering
  \includegraphics[width=\linewidth]{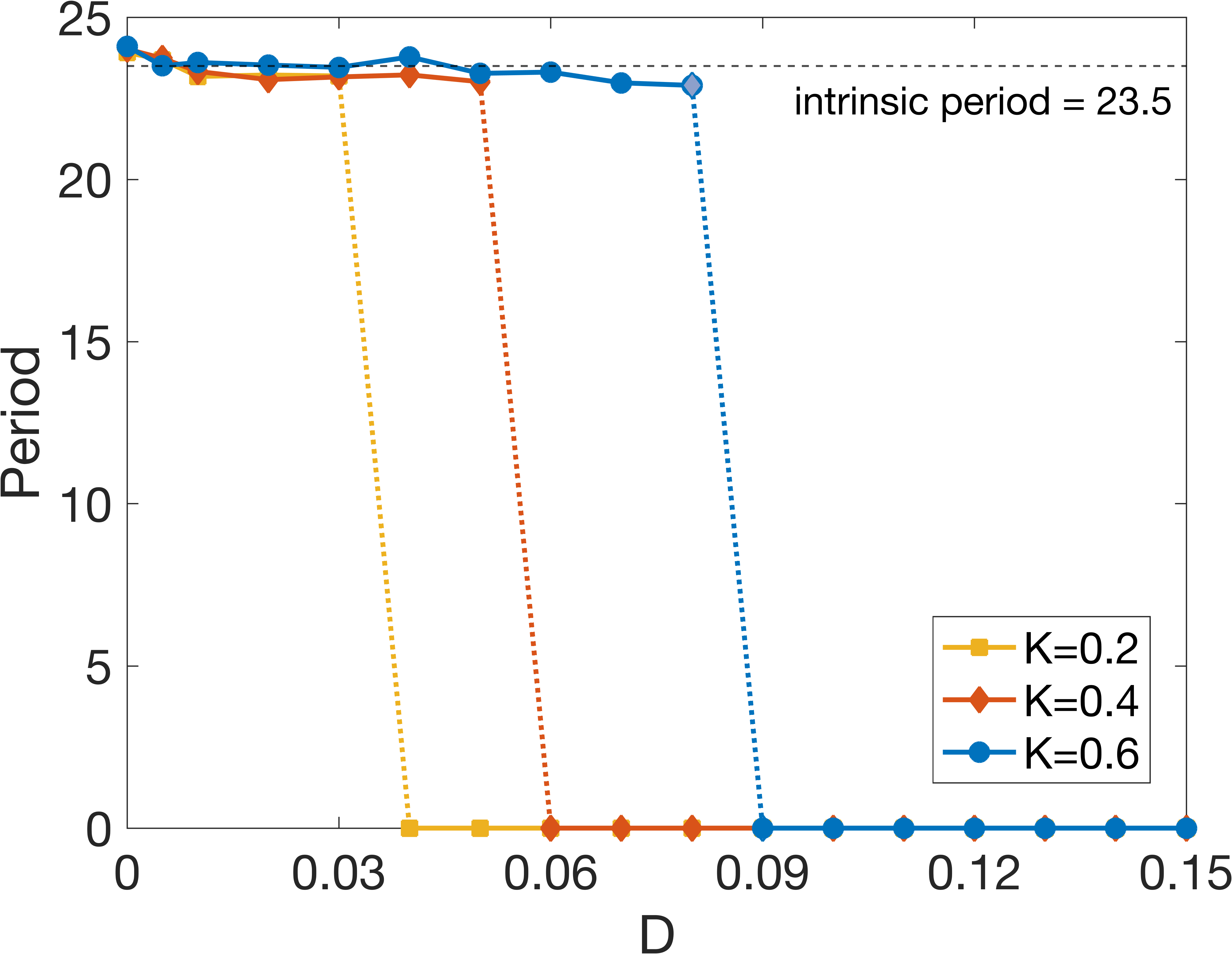}
  \caption{}
  \label{fig:periodD}
\end{subfigure}
    \caption{(a) Bifurcation diagram associated with the noise intensity ($D$). A Hopf bifurcation appears around the critical value $D_H \approx 0.8$ for $K$ fixed at $0.6$. Details of the figure are the same as those of Fig~\ref{fig:bifK2}. (b) Period of oscillations of $E[z]$ as a function of noise for different coupling strengths.}
    \label{fig:varyD}
\end{figure}

We argue that noise can affect synchrony-dependent rhythmicity. The noise can affect ensemble properties of oscillators including their coupling and their period of oscillations. In Fig~\ref{fig:jointPd}, we present solutions to (\ref{eq:mainPDE}) which are qualitatively different from the solutions shown in Fig~\ref{fig:jointPk}: as the intensity of the noise increases, the variance increases and the distribution tends to widen and shorten in all directions, indicating that a wider spread of values is possible and that external noise impairs the synchrony of the system. Although we showed that uniformly coupled networks can robustly synchronize (Fig~\ref{fig:jointPk}), it can also be concluded that noise weakens synchronization degree and affects the robustness of the system. Fig~\ref{fig:varyD} illustrates that the SCN is able to withstand higher noise levels with increasing coupling strength. However, noise eventually abrogate the oscillation in the SCN (Fig~\ref{fig:periodD}). For instance, when $K=0.6$, a noise intensity superior to $0.08$ is sufficient to desynchronize the neurons as indicated by the null period in Fig~\ref{fig:bifD}. When $D<0.08$, the system remains in synchrony with an ensemble period between $24.5$ and $22$. 

%-----------------------------------------
\section{Convergence studies} \label{sec:validation}
%-----------------------------------------
In this section, we report numerical results relating to the spatial accuracy of the numerical method. Our discussion focuses on a mesh convergence study to validate the order of convergence of the scheme in space. If the solution $\rho$ is sufficiently smooth, then the spatial discretization is expected to be second-order accurate (see Appendix \ref{app:method}). 

\begin{figure}[htbp]
    \centering
    \includegraphics[width=0.5\linewidth]{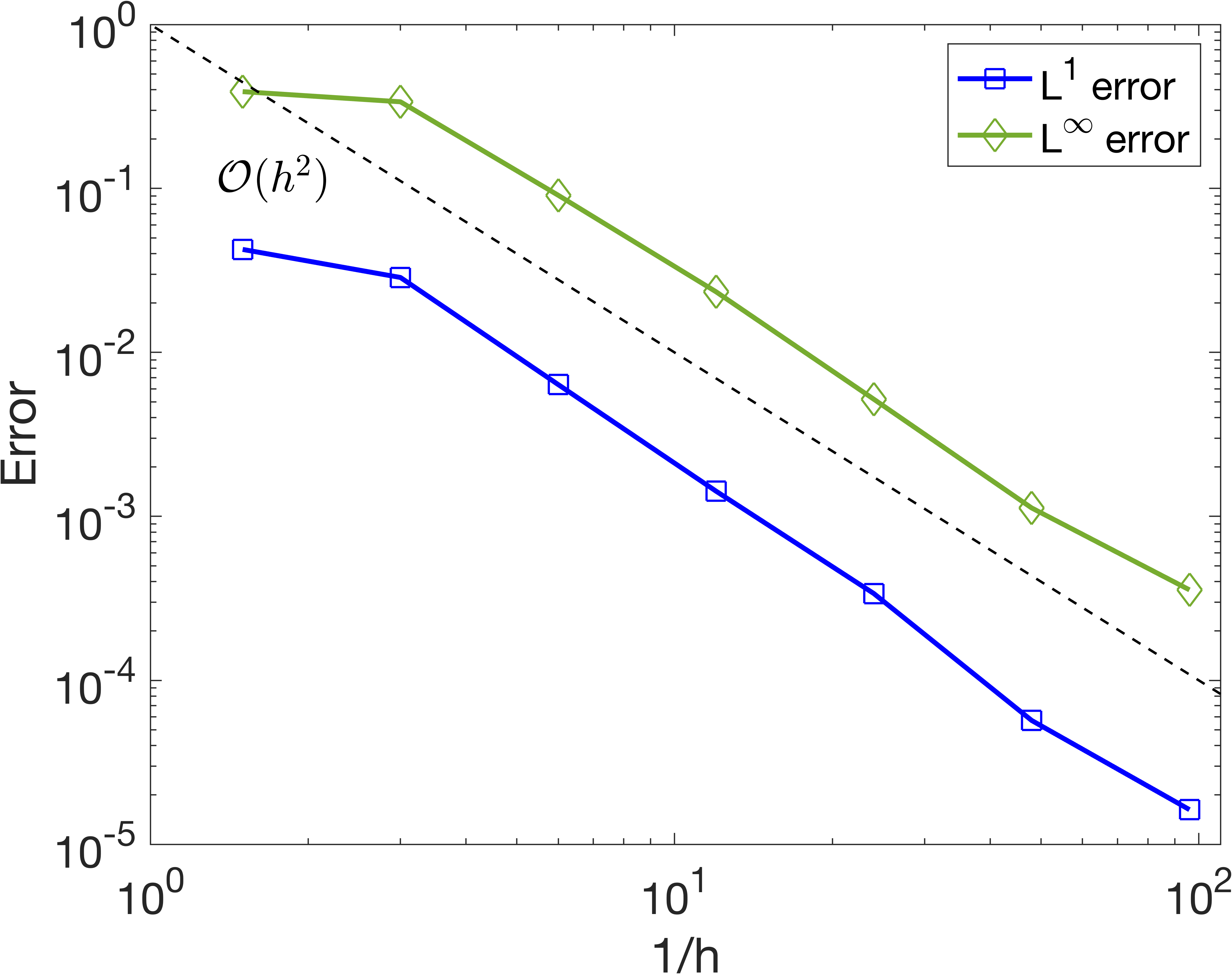}
    \caption{Convergence of error for the solution to the mean-field equation (\ref{eq:mainPDE}) in $L^1$ and $L^\infty$ norms. Grid cells are uniform in size across all three variables, $h = \Delta x = \Delta y = \Delta z$. The final time is t$_{final}$ = 1.}
    \label{fig:convError}
\end{figure}

\begin{figure}[htbp]
\centering
\begin{subfigure}{.48\textwidth}
  \centering
  \includegraphics[width=\linewidth]{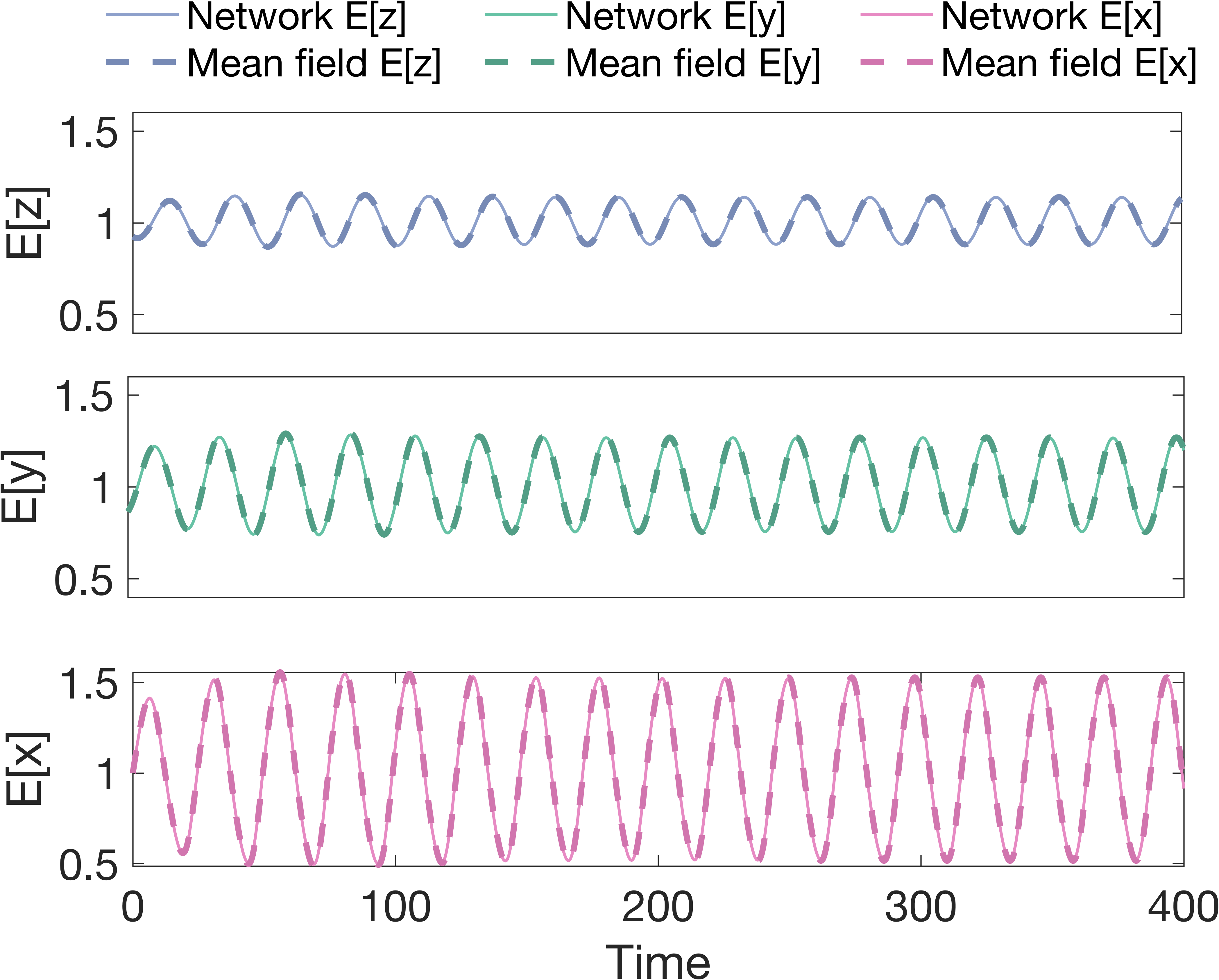} 
  \caption{}
  \label{fig:meanval}
\end{subfigure}
\hspace{0.2cm}
\begin{subfigure}{.48\textwidth}
  \centering
  \includegraphics[width=\linewidth]{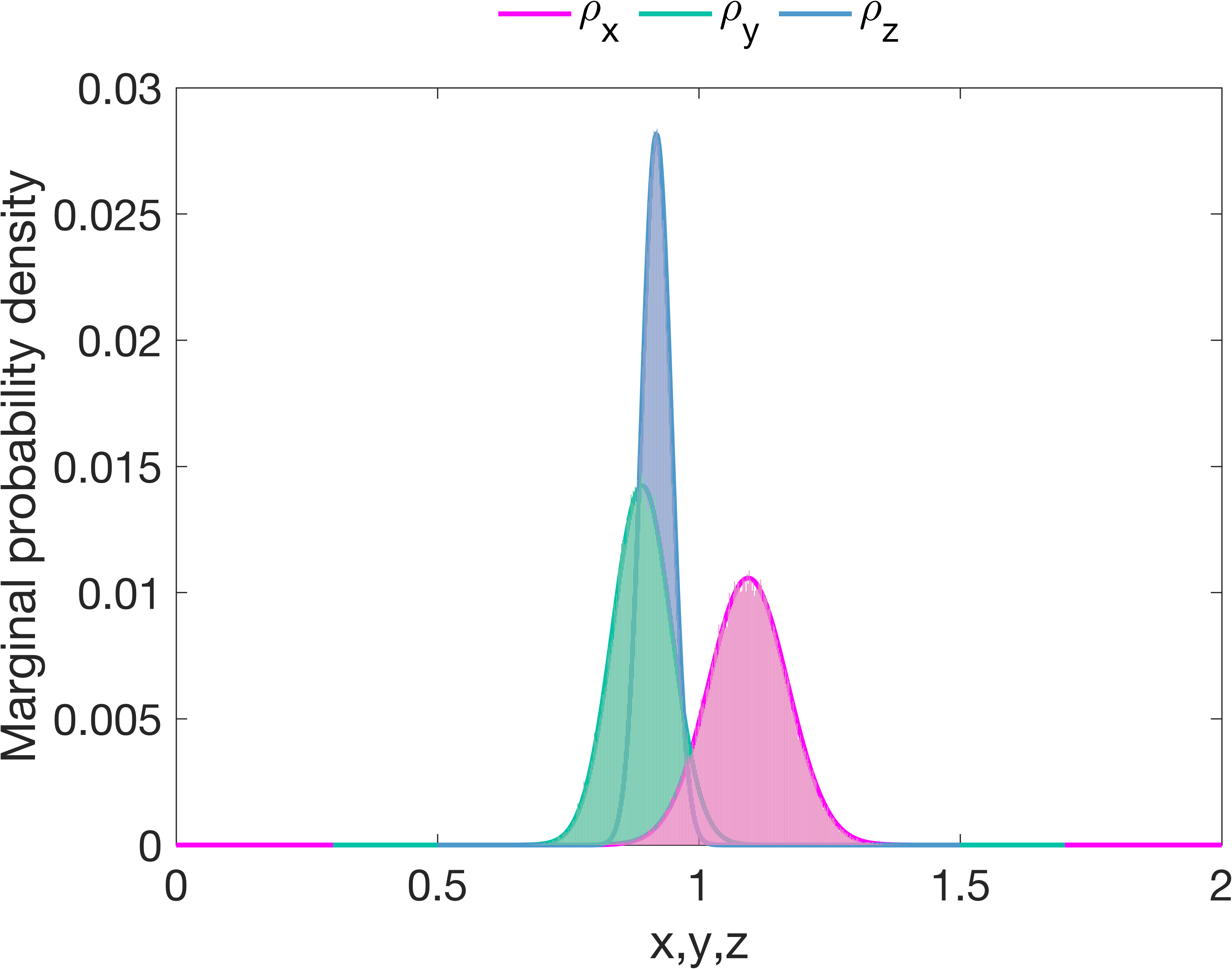} 
  \caption{}
  \label{fig:histval}
\end{subfigure}
\caption{(a) Time evolution of the averages in $x$, $y$, and $z$ obtained by simulating the network equations (solid curves) and the mean-field equation (dashed curves). We ran 100 Monte Carlo simulations of the network with network size N=100 up to time t$_{final}$ = 400. (b) Comparison between marginal probability densities $\rho_1(t,x)$, $\rho_2(t,y)$, $\rho_3(t,z)$ derived from the network and mean-field equation solutions. We conducted 10,000 Monte Carlo simulations with a network size N=10,000 up to time t$_{final}$ = 1.}
\label{fig:val-qual}
\end{figure}

In default of an analytical solution to our problem, we instead compute relative errors using different grid sizes. More precisely, we compute deviations from the estimated solution on a 3D fine mesh made of $384^3$ cells. Fig~\ref{fig:convError} illustrates the results in $L_1$ and $L_\infty$ norms. The time step $\Delta t$ is determined by the CFL condition derived in equation (\ref{eq:cfl}) and the spatial step size is uniform in all directions ($\Delta x$ = $\Delta y$ = $\Delta z$). We used zero-flux boundary conditions. The initial probability density function \review{$\rho(x, y, z, 0)$} is Gaussian, 
\begin{equation}
    \rho(x, y, z, 0) = \frac{1}{(2\pi)^{3/2}\sigma_{x_0}\sigma_{y_0}\sigma_{z_0}}e^{-(x-\mu_{x_0})^2 /(2\sigma_{x_0}^2) - (y-\mu_{y_0})^2 /(2\sigma_{y_0}^2) - (z-\mu_{z_0})^2 /(2\sigma_{z_0}^2)}.
\end{equation}
It appears from Fig~\ref{fig:convError} that the numerical scheme is at least second-order accurate, as anticipated. Fig~\ref{fig:val-qual} shows a qualitative similarity between solutions to the network equations and those obtained by solving the mean-field equation. In particular, we expect that the network and mean-field equation solutions will condense their mass around the periodic orbit \cite{baladron2012mean}. This is shown in Fig~\ref{fig:meanval} where the dynamics evolve to sustained oscillations. Table~\ref{tab:param-val} summarizes all of the parameters involved in the aforementioned simulations.

\begin{table}[htbp]
\centering
\begin{tabular}{lll}
\hline
Initial conditions     & Domain\quad        & Goodwin neuron        \\ \hline
$\mu_{x_0}=1$          & $x_{min}=0$   & $\alpha=1.8$                      \\
$\mu_{y_0}=0.9$        & $x_{max}=2$   & $n=20$                       \\
$\mu_{z_0}=0.9$        & $y_{min}=0$ & $K=0.5$                      \\
$\sigma_{x_0}^2=0.05$  & $y_{max}=2$ & $D=0.005$                    \\
$\sigma_{y_0}^2=0.02$  & $z_{min}=0$ & $\tau=6.4885$                    \\
$\sigma_{z_0}=0.01$   & $z_{max}=2$ &                                \\
& $\Delta x = 0.002$   &  \\ & $\Delta y = 0.002$ &   \\ & $\Delta z = 0.002$ &  \\ \hline
\end{tabular}
\caption{\label{tab:param-val} Model parameters. These parameters apply to the validation results presented in Section \ref{sec:validation}.}
\end{table}

%-----------------------------------------
\section{Discussion and conclusions} \label{sec:discussion}
%-----------------------------------------
To summarize, we have conducted an analysis of the Goodwin model using the mean-field limit approach from kinetic theory. We developed a minimal yet effective macroscopic model of the SCN circuit level dynamics and investigated the impact of noise on the emerging properties of the SCN --- rhythmicity (i.e., synchronisation), amplitude expansion, and ensemble period. We applied a positivity-preserving finite volume scheme developed in \cite{cch2015} for our numerical simulations. 

We presented simulation results indicating that coupling is important in maintaining the synchronization and amplitude expansion characteristics of the SCN, at least in the mean-field limit. Notably, increasing the coupling strength leads to phase transitions. We provided numerical evidences for the existence of Hopf bifurcations, with respect to the coupling parameter, which is synonymous with synchronized activity (Figs~\ref{fig:bifK2} and \ref{fig:jointPk}). On the one hand, low coupling strengths result in a decrease of the amplitude of the SCN rhythm. In particular, if the coupling strength is less than a certain threshold ($K_H$), the oscillation amplitude becomes null, meaning that the circadian rhythm is lost due to neuronal oscillators being out of phase with each other. Our findings, on the other hand, show that significant coupling causes resonance effects. This leads to amplitude expansion and the rapid establishment of a coherent evolution implying the dissipation of variance in the system.

Moreover, we provided a numerical description of the bifurcations that govern the instabilities caused by noise-induced transitions, i.e., Hopf bifurcations. Our approach allows us to identify where the system of coupled SCN neurons exhibit a stable stationary state (incoherence within the SCN network) or limit cycle oscillations (synchronized activity). We suggest that noise weakens synchrony-dependent rhythmicity and affects the robustness of the system (Fig~\ref{fig:jointPd}). Robustness to external noise decreases in proportion to the noise level: bifurcation boundaries are pushed forward as the noise level increases, making it more difficult to reach the oscillatory regime (Fig~\ref{fig:noise-bif}). However, in a biological context, rhythmicity could be recovered with higher activation rates ($k_1, k_3$), lower degradation rates ($k_2, k_4, k_6$) or slower inhibition of the mRNA by its inhibitor protein ($K_i$) in equation (\ref{eq:alpha}). 

\review{The repression mechanism used in modelling the negative feedback loop in circadian clocks can affect significantly properties of models, including robustness to perturbations. We use a Hill-type repression function to explain how transcriptional activity decreases as repressor concentration rises (see equation \ref{eq:repression}). Recently, a new mechanism of transcriptional repression based on protein sequestration has been proposed: repressors tightly bind activators to form an inactive 1:1 stoichiometric complex (see \cite{kim2012mechanism} for details). In Hill-type (HT) and protein-sequestration (PS) models, Kim and Forger investigated the qualitative differences based on the repression mechanisms \cite{kim2016protein,kim2012mechanism}.
According to their analyses, the HT and PS models have different prerequisites for generating rhythms: a large Hill exponent and a 1:1 molar ratio between repressor and activator, respectively. Kim and colleagues \cite{kim2014molecular} also showed that the coupled periods are near the mean period of the SCN when transcriptional repression occurs via protein sequestration, whereas the collective period is farther from the mean if modeled with Hill-type regulation. Apart from the repression mechanism, the models mentioned above are IBMs and differ from ours in that the coupling function is different, cells are heterogeneous in terms of period, and noise is not taken into account. In our mean-field model that uses Hill-type repression, we observe that the collective period is close to the intrinsic period of the cells in the presence of low to moderate noise. This could be explained by our use of a homogeneous network (see Fig~\ref{fig:perK2} and Fig~\ref{fig:periodD}). Moreover, according to Chen et al. \cite{chen2021collective}, there exist coupling strengths $c$ in both HT and PS models such that the collective frequency equals the average frequency of individual cells. For the HT model, such strength $c$ is larger.}

\review{Despite these differences between HT and PS models, many intercellular coupling properties are shared between the two and some general trends are similar. For instance, the logarithmic sensitivity of the repression function should be greater than 8 at steady state for both models to generate oscillations \cite{kim2016protein}, and increasing coupling strength causes amplitude expansion in both models \cite{kim2016protein,chen2021collective}. Our results can be extended when the protein sequestration function is used instead of the Hill function up to a constant in the bifurcation values for homogeneous networks of cells. Further study is needed when heterogeneous oscillators with different periods are coupled.}

\reviewtwo{In addition to nonlinearity in the repression function, oscillations require sufficiently long delays in feedback loops. This can be achieved by adding intermediate steps in the ODE formulation or by introducing explicit delays representing the durations of post-translational regulations. Our numerical scheme, first developed in \cite{cch2015}, applies when time-independent delays are modeled with noise, white or colored additive and multiplicative. However, numerical challenges may arise from adding explicit time delays. First, a delay may further constrain the stability condition on the time step so that the solver's time step is smaller than its value. Second, delays require storing a history of the function, which can be memory-prohibitive. Not the least is the noise effect, as noise causes the stochastic solution to disperse around the deterministic solution and the empirical variance stabilizes but for large time, with a limit value which increases as spatial regularity decreases and noise intensity increases. Importantly, the major numerical challenges would be due neither to noise nor to delay, but rather to the nature of the equation whose type degenerates at certain points of the domain of definition or at the boundary of this domain. The proposed scheme is able to cope with non-smooth stationary states, different time scales including metastability, as well as concentrations and self-similar behavior induced by singular nonlocal kernels \cite{cch2015}.}

\authors{Lastly}, our model has limitations, which should be acknowledged. It has been shown that the SCN is a heterogeneous network, consisting of two groups of neurons that are structurally and functionally different. Namely, the ventralateral part (VL) which receives light information and transmits it to the dorsalmedial part (DM). This second group is only indirectly sensitive to light \cite{noguchi2004clock}. Within these regions different neurotransmitters are used for communication between the cells \cite{gu2015noise}. Network topology, in addition to network heterogeneity, has a substantial influence on the SCN's collective behaviour. In this article, we tested an all-to-all \reviewtwo{linear} coupling between neurons, which may not be a realistic architecture for the SCN network. \reviewtwo{Extensions of our work could include a dual-network representation of the VL-DM architecture, as well as an emphasis on nonlinear cross-regional coupling.} \review{Future research could also look at the molecular details of the repression pathway, which could include both phosphorylation and protein sequestration.}

\appendix
%-----------------------------------------
\section{Convergence of stochastic and mean-field solutions}
%-----------------------------------------
We present, in this section, supplementary convergence results: the convergence of error between solutions to the stochastic system (\ref{eq:stochnetwork2}) and the mean-field equation (\ref{eq:mainPDE}) when the number of neurons tends to infinity. We have used a population of 10,000 Goodwin-type neurons and ran 10,000 Monte Carlo simulations of the network model using the Euler-Maruyama method \cite{bayram2018numerical}. Solutions of the network are constrained to remain in a smooth positive domain $D$. Namely, we simulate the case where the boundary $\partial D$ is instantaneously reflecting in an oblique direction. See \cite{gobet2001euler, hanks2017reflected} and the references therein. 
\begin{figure}[htbp]
\begin{subfigure}{.48\textwidth}
  \centering
  \includegraphics[width=\linewidth]{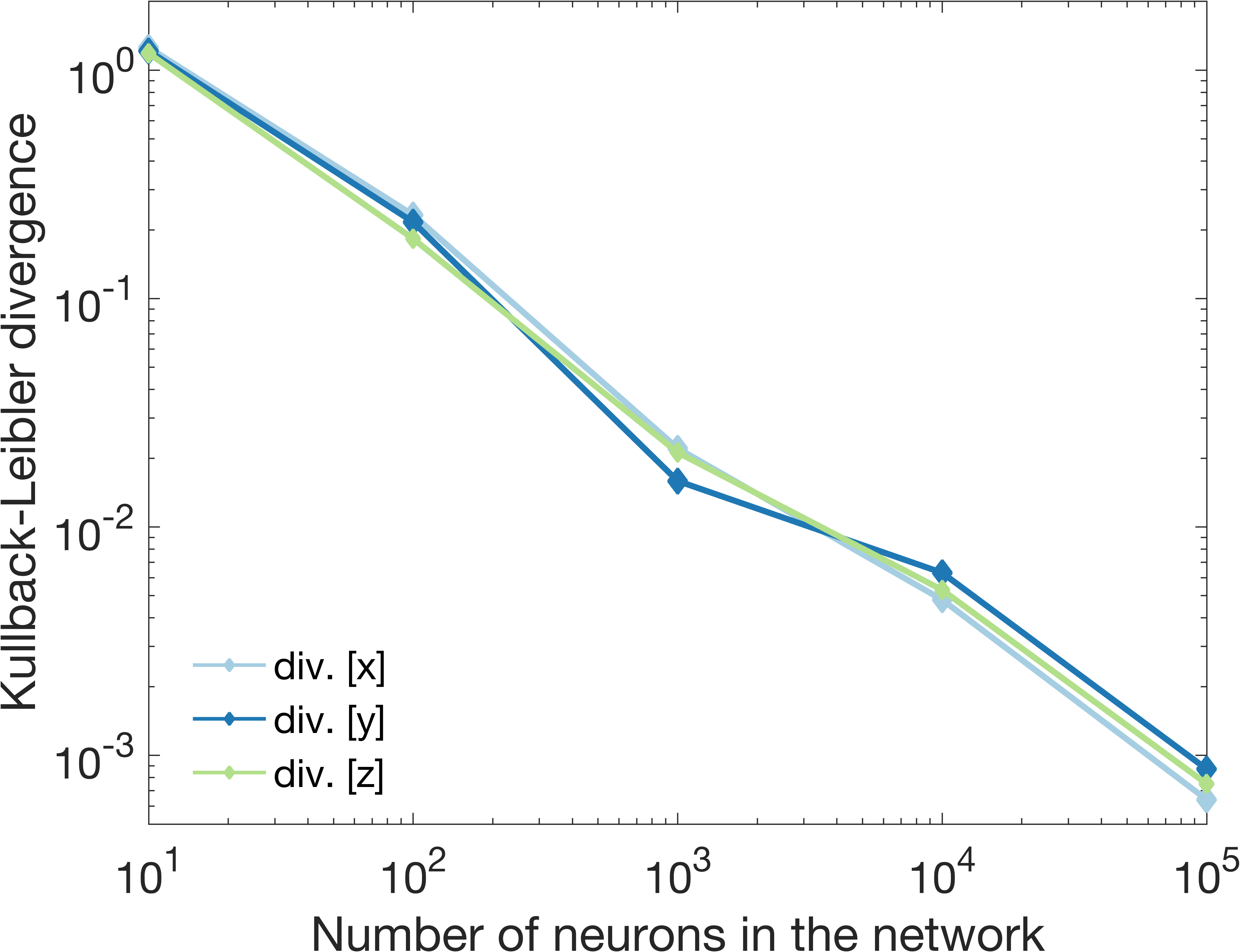} 
  \caption{}
  \label{fig:KL}
\end{subfigure}
\hspace{0.1cm}
\begin{subfigure}{.48\textwidth}
  \centering
  \includegraphics[width=\linewidth]{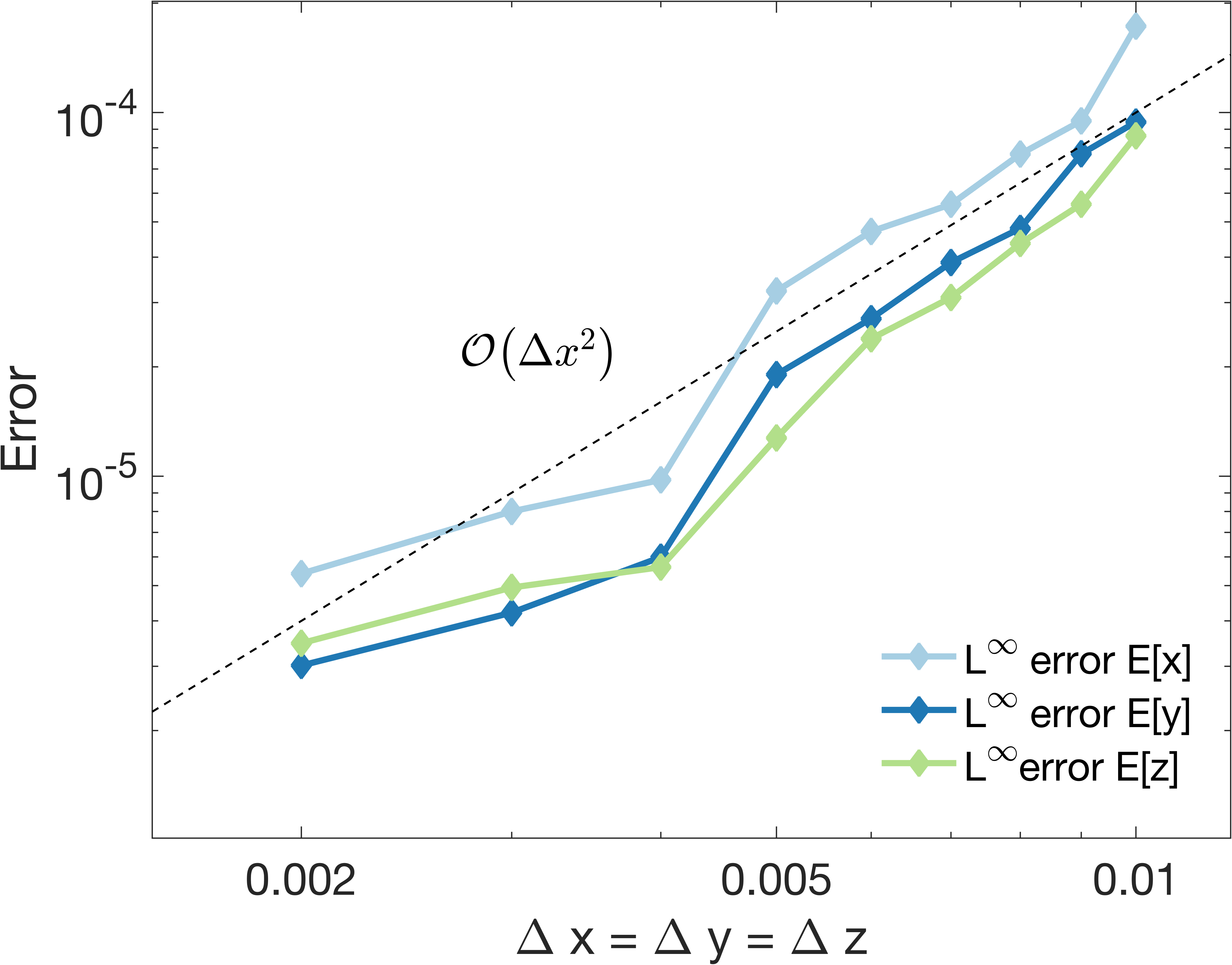}
  \caption{}
  \label{fig:Lerror}
\end{subfigure}
\caption{\review{(a) Kullback-Leibler divergence} between the marginal probability densities $\rho_1(t,x)$, $\rho_2(t,y)$, $\rho_3(t,z)$ calculated from the network and mean-field equation solutions as network size N increases. \review{(b) Convergence of error} between solutions to the stochastic system (\ref{eq:stochnetwork2}) and the mean-field equation (\ref{eq:mainPDE}) for the averages in $x$, $y$, and $z$ in $L_1$ and $L_\infty$ norms. Grid cells are assumed to be of uniform size in all three variables, $\Delta x = \Delta y = \Delta z$. We conducted 10,000 Monte Carlo simulations with a network size N=10,000 up to time t$_{final}$ = 1.}
\label{fig:val-quant}
\end{figure}

The most classical way to show convergence is to reason in terms of trajectories and to show that, when the number of agents tends to infinity, the behavior of the stochastic system converges to the mean-field approximation almost surely or in probability. Thus, we present the Kullback-Leibler divergence $D_{KL}\big(\rho^{Network}_{x,y,z}||\rho^{Mean-field}_{x,y,z}\big)$ between marginal distributions.

For increasing values of network size N, we ran 10,000 Monte Carlo simulations of the network equations until $t_{final} = 1$. As seen in Fig~\ref{fig:KL}, the Kullback-Leibler divergence decreases as N increases, validating the efficiency of the mean-field model even for relatively small values of N. We conclude that the solution to the mean-field equation (\ref{eq:mainPDE}) accurately represents the network's average behaviour. These results highlight the accuracy of the numerical method in preserving long time behavior of the solutions. Solution remain strictly positive for all $t > 0$, thus problem is not degenerate (see \ref{rem:positivity}). 

Next, consider the spatial averages of $x$, $y$, and $z$ separately, \textit{i.e.} $E[x]$, $E[y]$ and $E[z]$. An estimation of the relative error in $L_\infty$ norm at time $T$ is given by:
\begin{equation}
    e_{\Delta x}^\infty = \frac{\norm{\mu_{\Delta x}^{MF}(T) - \mu_{\Delta x}^{MC}(T)}_{L_\infty(\Omega)}}{\norm{\mu_{\Delta x}^{MC}(T)}_{L_\infty(\Omega)}},
\end{equation}
 where $\mu_{\Delta x}^{MF}$ represents the average in $x$ of the probability density computed on a uniform mesh of size $\Delta x$, and $\mu_{\Delta x}^{MC}$ represents the average in $x$ from the Monte Carlo simulations of the network equations using a similar mesh size. Relative errors $e_{\Delta y}^\infty$ and $e_{\Delta z}^\infty$ are computed similarly. Results are shown in Fig~\ref{fig:Lerror}. For reference, a dashed black line of slope two is added. We see that the slope of the dashed line appears to match well that of the error curves, suggesting that the mean-field equation accurately describes the network for large $N$.

%-----------------------------------------
\section{Derivation of the continuum model (\ref{eq:mainPDE})} \label{app:proof}
%-----------------------------------------
The equations in (\ref{eq:mainPDE}--\ref{eq:convol}) can be derived via the \textit{mean-field limit}. Here we present a simple formal description of this procedure. Starting from the deterministic model, define the empirical distribution density associated to a solution $(x(t), y(t), z(t))$ of (\ref{eq:network}) and given by
\begin{equation*}
\rho^N(x,y,z,t) = \frac{1}{N} \sum_{i=1}^{N} \delta(x-x_i(t))\delta(y-y_i(t))\delta(v-z_i(t)), \quad t>0,\label{eq:empirical}
\end{equation*}
where $\delta$ is the Dirac delta probability measure. Let us denote by $\mathcal{P}\big(\mathbb{R}^k\big)$ the space of probability measures on $\mathbb{R}^k$.

Let us assume that the particles remain in a fixed compact domain $(x_i(t), y_i(t),$\\ $z_i(t))\in\Bar{\Omega}\subset\mathbb{R}\times\mathbb{R}\times\mathbb{R}$ for all N in the time interval $t\in[0,T]$. Our model (\ref{eq:network}) satisfies this assumption if for instance the initial configuration is obtained as an approximation of an initial compactly supported probability measure $\rho_0$ \cite{carrillo2010particle}. Since for each $t$ the measure $\rho^N(t):=\rho^N(\cdot,\cdot,\cdot,t)$ is a probability measure in $\mathcal{P}\big(\mathbb{R}^{3}\big)$ with the uniform support in N, then Prohorov’s theorem implies that the sequence is weakly-$*$-relatively compact. Assume there exists a subsequence $\big(\rho^{N_k}\big)_k$ and $\rho:[0,T]\rightarrow\mathcal{P}\big(\mathbb{R}^{3}\big)$ such that
$
    \rho^{N_k}\rightarrow\rho\quad(k\rightarrow\infty)$ in the $ w^*\text{--convergence sense in } \mathcal{P}\big(\mathbb{R}^{3}\big),
$
pointwise in time. Following the approach in \cite{carrillo2010particle}, let us consider the test function $\varphi \in C_0^1\big(\mathbb{R}^{3}\big)$. To simplify the notation we will write $\varphi$ for $\varphi \big(x_i(t), y_i(t), z_i(t)\big)$, and $x_i$, $y_i$, and $z_i$ for $x_i(t)$, $y_i(t)$, and $z_i(t)$, respectively. We compute
\begin{align*}
    \dv{}{t} &\Big\langle\rho^N(t), \varphi \Big \rangle = \frac{1}{N} \sum_{i=1}^{N} \dv{}{t} \varphi \big(x_i(t), y_i(t), z_i(t)\big) \\
        &= \frac{1}{N} \sum_{i=1}^{N} \partial_x \varphi \, \Big(\frac{\alpha}{1+z_i^n} - x_i + \frac{K}{N}\sum_{j=1}^{N}H(x_i-x_j) \Big)+ \frac{1}{N} \sum_{i=1}^{N} \partial_y \varphi  \, (x_i-y_i) \\
        &\hspace{0.5cm}+ \frac{1}{N} \sum_{i=1}^{N} \partial_z \varphi  \, (y_i-z_i) \\
        &=\big\langle\rho^N(t),\partial_y\varphi \,(x-y)\big\rangle+\big\langle\rho^N(t),\partial_z\varphi \,(y-z)\big\rangle\\
        &\hspace{1cm}+\Bigg(\frac{1}{N}\sum_{i=1}^{N}\partial_x\varphi \,\Big(\frac{\alpha}{1+z_i^n}-x_i\Big)\Bigg)+ \Bigg(\frac{1}{N}\sum_{i=1}^{N}\bigg[\frac{K}{N}\sum_{j=1}^{N}H(x_i-x_j)\bigg] \,\partial_x\varphi\Bigg)\\
        &=\big\langle\rho^N(t),\partial_y\varphi \,(x-y)\big\rangle+\big\langle\rho^N(t),\partial_z\varphi \,(y-z)\big\rangle\\
        &\hspace{1cm}+\bigg\langle\rho^N(t),\Bigg(\frac{\alpha}{1+z^n}-x\Bigg) \,\partial_x\varphi\bigg\rangle+\bigg\langle\rho^N(t),\Bigg(\frac{K}{N}\sum_{j=1}^{N}H(x-x_j)\Bigg) \,\partial_x\varphi\bigg\rangle.
\end{align*}
We can rewrite 
\begin{equation*}
    \frac{1}{N}\sum_{j=1}^{N}H(x-x_j)=\frac{1}{N}\sum_{j=1}^{N}\big\langle H(x-\omega),\delta(\omega-x_j)\big\rangle_x=H\star m_{\rho^N}(y,z,t),
\end{equation*}
where
\begin{equation*}
    m_{\rho^N}(y,z,t) =  \int_{\mathbb{R}} \rho^N(x,y,z,t) dx = \Bigg\langle 1,\frac{1}{N} \sum_{j=1}^{N} \delta(\omega-x_j)\delta(y-y_j)\delta(z-z_j) \Bigg\rangle_x ;
\end{equation*}
Collecting all the terms we obtain
\begin{equation*}
    \dv{}{t} \Big\langle \rho^N(t), \varphi \Big \rangle 
        = \Big\langle \rho^N(t), \partial_x \varphi  \, \Big(\frac{\alpha}{1+z^n} - x + K \big(H \star m_{\rho^N}\big)\Big) + \partial_y \varphi  \, (x-y) + \partial_z \varphi  \, (y-z) \Big \rangle .
\end{equation*}
After integration by part in $x,y,$ and $z$, we obtain
\begin{equation*}
\Bigg\langle \pdv{f}{t} + \partial_x  \, \Big[ \xi\big(\rho^N\big)\rho^N\Big] + \partial_y  \, \Big[(x-y) \rho^N\Big] + \partial_z  \, \Big[(y-z) \rho^N\Big], \varphi \Bigg\rangle = 0
\end{equation*}
or, in the strong form,
\begin{equation*}
\pdv{f}{t} + \partial_x  \, \Big[ \xi\big(\rho^N\big)\rho^N\Big] + \partial_y  \, \Big[(x-y) \rho^N\Big] + \partial_z  \, \Big[(y-z) \rho^N\Big] = 0,
\end{equation*}
where $\xi$ is defined by 
\begin{equation*}
    \xi\big(\rho\big) \big(x,y,z,t\big) = \frac{\alpha}{1+z^n} - x +  K \big(H\star\rho\big),
\end{equation*}
with 
\begin{equation*}
    H \star \rho(x,y,z,t) = \int_{\mathbb{R}^{3d}} H(x-w) \rho(w,y,z,t) \,dw\,dy\,dz \,.
\end{equation*}
Letting $k \rightarrow \infty$ in the subsequence $\rho^{N_k}$ leads formally to
\begin{equation*}
   \pdv{\rho}{t} - \partial_x \, \big[\xi(\rho)\rho\big] -  \partial_y \, \big[(x-y) \rho \big] -  \partial_z \, \big[(y-z)\rho\big] = 0.
\end{equation*}
The case with noise in \eqref{eq:mainPDE} follows a similar approach using the so-called coupling method introduced by Sznitman \cite{sznitman1991topics} together with \cite{Lions-Sznitman-1984-SDEreflectingBC,Sznitman-1984-nonlinear-reflecting-diffusion} to deal with boundary conditions. Defining a system of uncoupled copies of McKean-Vlasov particles and comparing the error with respect to the coupled system of particles is a common approach in many areas of applications of interacting particle systems in mathematical biology, see \cite{carrillo2010particle} for instance. By taking the difference between the two particle systems, one can develop direct Gronwall inequalities for the 2-Wasserstein distance among the marginals of the joint probability distributions. We refer the reader for the details to \cite{bolley2011stochastic} for instance.

%-----------------------------------------
\section{Presentation of the numerical scheme} \label{app:method}
%-----------------------------------------
In this section, we present our finite volume scheme for (\ref{eq:mainPDE}) preserving the structure of the gradient flow in the case of identical oscillators. We also prove the positivity preserving property for this scheme.

Inspired by \cite{cch2015, carrillo2019kuramoto,KRUK2021110275}, we construct a discrete numerical scheme in the variables $x$, $y$ and $z$ in (\ref{eq:mainPDE}) as follows. We introduce a Cartesian mesh consisting of the cells  $C_{i,j,k} := \big[x_{i-\frac{1}{2}}, x_{i+\frac{1}{2}}\big] \times \big[y_{j-\frac{1}{2}}, y_{j+\frac{1}{2}}\big] \times \big[z_{k-\frac{1}{2}}, z_{k+\frac{1}{2}}\big]$, which for the sake of simplicity are assumed to be of uniform size $\Delta x \Delta y \Delta z$, that is, $x_{i+\frac{1}{2}} - x_{i-\frac{1}{2}} \equiv \Delta x$, $\forall$ $i$, $y_{j+\frac{1}{2}} - y_{j-\frac{1}{2}} \equiv \Delta y$, $\forall$ $j$, and $z_{k+\frac{1}{2}} - z_{k-\frac{1}{2}} \equiv \Delta z$, $\forall$ $k$. 

Here, we denote by 
\begin{equation}
    \Bar{\rho}_{i,j,k}(t) = \frac{1}{\Delta x\Delta y\Delta z}
    \iiint_{C_{i,j,k}} \rho(x,y,z,t) \,dx\,dy\,dz 
\end{equation}
the computed cell averages of the solution $\rho$, which we assume to be known or approximated at time $t \geq 0$. A discrete finite volume scheme is obtained by integrating (\ref{eq:mainPDE}) over each cell $C_{i,j,k}$ and is given by the following system of ODEs for $\Bar{\rho}_{i,j,k}$:

\begin{align}
    \dv{\Bar{\rho}_{i,j,k}(t)}{t} = -\frac{F^x_{i+\frac{1}{2},j,k}(t)-F^x_{i-\frac{1}{2},j,k}(t)}{\Delta x} - & \frac{F^y_{i,j+\frac{1}{2},k}(t)-F^y_{i,j-\frac{1}{2},k}(t)}{\Delta y} \nonumber \\ - & \frac{F^z_{i,j,k+\frac{1}{2}}(t)-F^z_{i,j,k-\frac{1}{2}}(t)}{\Delta z},
    \label{eq:odes}
\end{align}
where $F^x_{i+\frac{1}{2},j,k}$, $F^y_{i,j+\frac{1}{2},k}$ and $F^z_{i,j,k+\frac{1}{2}}$ are upwind numerical fluxes and approximate the continuous fluxes in the $x$, $y$ and $z$ directions, respectively. For simplicity, we will omit the dependence of the computed quantities on $t \geq 0$. In order to construct the upwind fluxes, we first construct piecewise linear polynomials in each cell $C_{i,j,k}$,
\begin{equation}
\begin{aligned}
    \Tilde{\rho}_{i,j,k}(x,y,z) = \Bar{\rho}_{i,j,k} + (\rho_x)_{i,j,k} &(x-x_i) + (\rho_y)_{i,j,k} (y-y_j) \\ &+ (\rho_z)_{i,j,k} (z-z_k), \quad (x,y,z) \in C_{i,j,k}
    \label{eq:polyreconstruction}
\end{aligned}
\end{equation}
and compute the right (“east”),  $\rho_{i,j,k}^E$, and left (“west”),  $\rho_{i,j,k}^W$, point values at the corresponding cell interfaces $(x_{i+\frac{1}{2}},y_j,z_k)$, $(x_{i-\frac{1}{2}},y_j,z_k)$, $(x_i,y_{j+\frac{1}{2}},z_k)$, $(x_i,y_{j-\frac{1}{2}},z_k)$, $(x_i,y_j,z_{k+\frac{1}{2}})$ and $(x_i,y_j,z_{k+\frac{1}{2}})$. Namely,
\begin{align}
    \rho_{i,j,k}^{E_x} &= \Tilde{\rho}_{i,j,k}(x_{i+\frac{1}{2}}-0,y_j,z_k) = \Bar{\rho}_{i,j,k} + \frac{\Delta x}{2} (\rho_x)_{i,j,k}, \nonumber \\
    \rho_{i,j,k}^{W_x} &= \Tilde{\rho}_{i,j,k}(x_{i-\frac{1}{2}}+0,y_j,z_k) = \Bar{\rho}_{i,j,k} - \frac{\Delta x}{2} (\rho_x)_{i,j,k}.
    \label{eq:pval-x}
\end{align}
and analogously for the other two variables.

These values will be second-order accurate provided the numerical derivatives $(\rho_x)_{i,j,k}$, $(\rho_y)_{i,j,k}$ and $(\rho_z)_{i,j,k}$ are at least first-order accurate approximations. To ensure the point values in (\ref{eq:pval-x}) are both second-order and nonnegative, the slopes $(\rho_x)_{i,j,k}$, $(\rho_y)_{i,j,k}$, $(\rho_z)_{i,j,k}$ are calculated according to the following adaptive procedure. First, the centered-difference approximations
\begin{align}
    (\rho_x)_{i,j,k} = \frac{\rho_{i+1,j,k}-\rho_{i-1,j,k}}{2\Delta x}, &\quad
    (\rho_y)_{i,j,k} =
    \frac{\rho_{i,j+1,k}-\rho_{i,j-1,k}}{2\Delta y} \nonumber \\
    \text{and} \quad
    (\rho_z)_{i,j,k} &= \frac{\rho_{i,j,k+1}-\rho_{i,j,k-1}}{2\Delta z}
\end{align}
are used for all $i,j,k$. Then, if the reconstructed point values in some cell $C_{i,j,k}$ become negative (i.e., either $\rho^E_{i,j,k}<0$ or $\rho^W_{i,j,k}<0$), we recalculate the corresponding slopes $(\rho_x)_{i,j,k}$, $(\rho_y)_{i,j,k}$ or $(\rho_z)_{i,j,k}$ using a monotone nonlinear slope limiter, which guarantees that the reconstructed point values are nonnegative as long as the cell averages $\Bar{\rho}_{i,j,k}$ are nonnegative for all $i,j,k$. In our numerical experiments, we have used the one-parameter family of the generalized minmod limiter \cite{cch2015,lie2003,nessyahu1990,sweby1984,vanleer1979}: 
\begin{align}
     (\rho_x)_{i,j,k} &= \text{minmod}\Big( \theta \frac{\Bar{\rho}_{i+1,j,k}-\Bar{\rho}_{i,j,k}}{\Delta x}, \frac{\Bar{\rho}_{i+1,j,k}-\Bar{\rho}_{i-1,j,k}}{2\Delta x}, \theta \frac{\Bar{\rho}_{i,j,k}-\Bar{\rho}_{i-1,j,k}}{\Delta x}\Big) 
    \label{eq:minmod}
\end{align}
and analogously for the other two variables, where the minmod function and its parameters are chosen as in \cite{cch2015}.

Given the polynomial reconstruction (\ref{eq:polyreconstruction}) and its point values (\ref{eq:pval-x}), the upwind numerical fluxes in (\ref{eq:odes}) are defined as
\begin{align}
    F^x_{i+\frac{1}{2},j,k} &= \xi_{i+\frac{1}{2},j,k}^{+}  \rho_{i,j,k}^{E_x} + \xi_{i+\frac{1}{2},j,k}^{-}  \rho_{i+1,j,k}^{W_x}  \nonumber \\
    F^y_{i,j+\frac{1}{2},k} &= u_{i,j+\frac{1}{2},k}^{+}  \rho_{i,j,k}^{E_y} + u_{i,j+\frac{1}{2},k}^{-}  \rho_{i,j+1,k}^{W_y} \nonumber \\
    F^z_{i,j,k+\frac{1}{2}} &= v_{i,j,k+\frac{1}{2}}^{+}  \rho_{i,j,k}^{E_z} + v_{i,j,k+\frac{1}{2}}^{-}  \rho_{i,j,k+1}^{W_z},
    \label{eq:fluxes}
\end{align}
where the discrete values $\xi_{i+\frac{1}{2},j,k}$, $u_{i,j+\frac{1}{2},k}$ and $v_{i,j,k+\frac{1}{2}}$ of the velocities at midpoints are obtained as follows,
\begin{equation}
\begin{aligned}
    \xi_{i+\frac{1}{2},j,k} &= - \Bigg(\frac{D}{\Delta x} \log \frac{\Bar{\rho}_{i+1,j,k}}{\Bar{\rho}_{i,j,k}} - \frac{f^x_{i,j,k} + f^x_{i+1,j,k}}{2} - K \Big(\Delta x \Delta y \Delta z \sum_{i,j,k} \mathbf{x} \Bar{\rho}_{i,j,k}  - x_{i+\frac{1}{2}} \Big) \Bigg) \\
    u_{i,j+\frac{1}{2},k} &= \frac{f^y_{i,j,k} + f^y_{i,j+1,k}}{2} , \quad v_{i,j,k+\frac{1}{2}}= \frac{f^z_{i,j,k} + f^z_{i,j,k+1}}{2},
    \label{eq:velocities}
\end{aligned}
\end{equation}
and the positive and negative parts are denoted by
\begin{align}
    \xi_{i+\frac{1}{2},j,k}^{+} = \max\big(\xi_{i+\frac{1}{2},j,k},0\big), \quad \xi_{i+\frac{1}{2},j,k}^{-} = \min\big(\xi_{i+\frac{1}{2},j,k},0\big) 
    \label{eq:minmax}
\end{align}
and analogously for the other two variables. We note that  $\mathbf{x}=[x_{\frac{1}{2}}, x_{1+\frac{1}{2}}, \dots, x_{N+\frac{1}{2}}]$ in (\ref{eq:velocities}) is a row vector of  (inter)face values of the cells in the $x$-direction, and the values $f^x_{i,j,k}$, $f^y_{i,j,k}$, $f^z_{i,j,k}$ are calculated by discretizing (\ref{eq:smallf}):
\begin{align}
    f^x(x,y,z) := \frac{\alpha}{1+z^n} - x, \quad 
    f^y(x,y,z) := x - y, \quad
    f^z(x,y,z) := y - z.
    \label{eq:smallf}
\end{align}

Finally, the semi-discrete scheme (\ref{eq:odes}) is integrated using a stable and accurate ODE solver. In all our numerical examples, the third-order strong preserving Runge-Kutta (SSP-RK) ODE solver \cite{gottlieb2001ssp} is used.

\begin{remark}
    The second-order finite volume scheme (\ref{eq:odes}),(\ref{eq:fluxes})--(\ref{eq:minmax}), reduces to the first-order scheme if the piecewise constant reconstruction is used instead of (\ref{eq:polyreconstruction}), in which case we have
    $
        \Tilde{\rho}_{i,j,k}(x,y,z) = \Bar{\rho}_{i,j,k}  
    $
    and therefore
    \[ \rho_{i,j,k}^{E_x}=\rho_{i,j,k}^{W_x}=\rho_{i,j,k}^{E_y}=\rho_{i,j,k}^{W_y}=\rho_{i,j,k}^{E_z}=\rho_{i,j,k}^{W_z}= \Bar{\rho}_{i,j,k}, \quad \forall i,j,k.
    \]
\end{remark}

\begin{remark} \label{rem:positivity}
Given initial data $\rho_0(x) \geq 0$ for system (\ref{eq:mainPDE}), the semi-discrete finite-volume scheme (\ref{eq:odes}),(\ref{eq:fluxes})--(\ref{eq:minmax}) preserves positivity for all $t>0$. A CFL condition can be computed explicitly using equation (\ref{eq:odes}) which is discretized by the forward Euler method. Specifically, the computed cell averages $\Bar{\rho}_{i,j,k} \geq 0$, $\forall$ $i,j,k$ provided that the following CFL condition is satisfied:
\begin{align}
    \Delta t \leq \min \Bigg\{\frac{\Delta x}{6a}, \frac{\Delta y}{6b}, \frac{\Delta z}{6c} \Bigg\}, \quad \text{where} \quad
    a &= \max_{i,j,k} \Bigg\{\xi_{i+\frac{1}{2},j,k}^{+}, -\xi_{i-\frac{1}{2},j,k}^{-}\Bigg\}, \nonumber\\
    b = \max_{i,j,k} \Bigg\{u_{i,j+\frac{1}{2},k}^{+}, -u_{i,j-\frac{1}{2},k}^{-}\Bigg\}, \quad 
    c &= \max_{i,j,k} \Bigg\{v_{i,j,k+\frac{1}{2}}^{+}, -v_{i,j,k-\frac{1}{2}}^{-}\Bigg\}, 
    \label{eq:cfl}
\end{align}
with $\xi_{i+\frac{1}{2},j,k}^{\pm}$, $u_{i,j+\frac{1}{2},k}^{\pm}$ and $v_{i,j,k+\frac{1}{2}}^{\pm}$ defined in (\ref{eq:minmax}).
\end{remark}

\begin{remark}\label{sec:GPU}
{\bf Numerical simulations with GPUs.-}
The finite volume algorithm for solving the mean-field equation described in Appendix \ref{app:method} is computationally very expensive. In fact, when the discretization steps $\Delta x$, $\Delta y$, and $\Delta z$ are small, we must also maintain $\Delta t$ small enough to ensure the algorithm's stability (see \ref{rem:positivity}). The simulations will undoubtedly slow down as a result of this. We were able to mitigate this issue by employing more powerful hardware, specifically graphical processing units (GPUs). Through GPU computing we were able to adopt a more accurate and stable ODE solver, namely the strong stability-preserving Runge-Kutta (SSP-RK) solver of order three \cite{gottlieb2001ssp}, thus allowing for three calls per time step at a lower computational cost.
\end{remark}

\section*{Acknowledgments}
The authors would like to thank the anonymous referees for valuable suggestions and remarks.

%\bibliographystyle{siamplain}
%\bibliography{references}

\end{document}

%% file: ex_shared.tex
\usepackage{lipsum}
\usepackage{amsfonts}
\usepackage{graphicx}
\usepackage{epstopdf}
\usepackage{algorithmic}
\ifpdf
  \DeclareGraphicsExtensions{.eps,.pdf,.png,.jpg}
\else
  \DeclareGraphicsExtensions{.eps}
\fi

\usepackage{physics}
\usepackage{subcaption}
\NewDocumentCommand{\evalat}{sO{\big}mm}{%
  \IfBooleanTF{#1}
   {\mleft. #3 \mright|_{#4}}
   {#3#2|_{#4}}%
}

\usepackage{enumitem}
\setlist[enumerate]{leftmargin=.5in}
\setlist[itemize]{leftmargin=.5in}

\newsiamremark{remark}{Remark}
\newsiamremark{hypothesis}{Hypothesis}
\crefname{hypothesis}{Hypothesis}{Hypotheses}
\newsiamthm{claim}{Claim}

\headers{Clocks ticking despite the noise}{S. M. C. Abo, J. A. Carrillo, and A. T. Layton}

\title{Can the clocks tick together despite the noise? Stochastic simulations and analysis \thanks{Submitted to the editors DATE.
\funding{ATL is supported by the Canada 150 Research Chairs Program and the Natural Sciences and Engineering Research Council of Canada (NSERC Discovery award: RGPIN-2019-03916). JAC is supported by the Advanced Grant Nonlocal-CPD (Nonlocal PDEs for Complex Particle Dynamics: Phase Transitions, Patterns and Synchronization) of the European Research Council Executive Agency (ERC) under the European Union's Horizon 2020 research and innovation programme (grant agreement No. 883363).
}}}

\author{Dianne Doe\thanks{Imagination Corp., Chicago, IL 
  (\email{ddoe@imag.com}, \url{http://www.imag.com/\string~ddoe/}).}
\and Paul T. Frank\thanks{Department of Applied Mathematics, Fictional University, Boise, ID 
  (\email{ptfrank@fictional.edu}, \email{jesmith@fictional.edu}).}
\and Jane E. Smith\footnotemark[3]}

\author{St\'ephanie M.C. Abo
    \thanks{Department of Applied Mathematics, University of Waterloo, Waterloo, Ontario, Canada
    (\email{sabo@uwaterloo.ca}).}
\and Jos\'e A. Carrillo 
    \thanks{Mathematical Institute, University of Oxford, Oxford OX2 6GG, UK
    (\email{carrillo@maths.ox.ac.uk}).}
\and Anita T. Layton 
    \thanks{Department of Applied Mathematics, Cheriton School of Computer Science, Department of Biology, and School of Pharmacy, University of Waterloo, Waterloo, Ontario, Canada
    (\email{anita.layton@uwaterloo.ca}).}
    }

\usepackage{amsopn}